\documentclass[superscriptaddress,prb,aps,twocolumn]{revtex4}
\usepackage{graphicx}
\usepackage{subfigure}
\usepackage{bm}
\usepackage{amsmath}
\begin{document} 
\title{Analyzing intramolecular vibrational energy redistribution via the
overlap intensity-level velocity correlator}
\author{Srihari Keshavamurthy}
\affiliation{Department of Physics, Washington State University, Pullman, WA
99164-2814}
\affiliation{Department of Chemistry, Indian Institute
of Technology, Kanpur, India 208 016}
\author{Nicholas R. Cerruti}
\affiliation{Department of Physics, Washington State University, Pullman, WA
99164-2814}
\author{Steven Tomsovic}
\affiliation{Department of Physics, Washington State University, Pullman, WA
99164-2814}
\date{\today}

\begin{abstract}
Numerous experimental and theoretical studies have established that
intramolecular vibrational energy redistribution (IVR) in isolated
molecules has a hierarchical tier structure. The tier structure implies
strong correlations between the energy level motions of a quantum system
and its intensity-weighted spectrum.  A measure, which explicitly accounts
for this correlation, was first introduced by one of us as a sensitive
probe of phase space localization.  It correlates eigenlevel velocities
with the overlap intensities between the eigenstates and some localized
state of interest.  A semiclassical theory for the correlation is
developed for systems that are classically integrable and complements
earlier work focusing exclusively on the chaotic case.  Application to a
model two dimensional effective spectroscopic Hamiltonian shows that the
correlation measure can provide information about the terms in the
molecular Hamiltonian which play an important role in an energy range of
interest and the character of the dynamics.  Moreover, the correlation
function is capable of highlighting relevant phase space structures
including the local resonance features associated with a specific bright
state.  In addition to being ideally suited for multidimensional systems
with a large  density of states, the measure can also be used to gain
insights into phase space transport and localization.  It is argued that
the overlap intensity-level velocity correlation function provides a
novel way of studying vibrational energy redistribution in isolated
molecules.  The correlation function is ideally suited to analyzing the
parametric spectra of molecules in external fields.
\end{abstract}
\maketitle

\nopagebreak

\section{Introduction}

The nature of the dynamics of a molecule in a highly excited rovibrational
state has been a subject of considerable study and debate for a number
of years\cite{rev}.  Deciding the fate of a localized excitation in a
molecule at high energies in terms of the time scales, pathways, and final
destination has posed a significant challenge for
experimentalists\cite{rev1,rev3,rev4,rev5, rev6,rev7,rev8} 
and theorists\cite{rev,rev2,rev9,rev10} alike.  Intramolecular vibrational
energy redistribution (IVR), as it has been called, is perhaps one of the
most important phenomenon in chemical physics.  It is now well established
that IVR plays a crucial role in our understanding of reaction rate
theories and the ability to effect mode specific chemistry.  The advent of
modern high resolution spectroscopic techniques\cite{rev3,rev6,rev7,rev8}
has led to a fairly detailed study of a number of molecules in highly
excited energy regions. Such beautiful and precise studies have
demonstrated the richness and complexity of IVR in various molecules and
are an essential step towards realizing the goal of quantum control of
chemical reactions. 

As a result of a number of experimental and theoretical studies, a
standard paradigm for IVR called the tier model has
evolved\cite{rev,rev1,rev2,rev3,rev4,rev5,rev6,rev7,rev8}.
In this model an optically bright state is coupled to a bath of
dark states in a hierarchical fashion.  In a typical experimental
situation, the bright state corresponds to the eigenstate of the
zeroth-order molecular Hamiltonian $H_{0}$ which carries almost all of
the  oscillator strength.  The form for this zeroth-order Hamiltonian
involves the rigid rotor-harmonic oscillator part corrected 
perturbatively for nonresonant anharmonic interactions.  The rest of the
eigenstates of $H_{0}$ having negligible oscillator strength are termed
dark states. At low energies the excitation is localized since the
bright state is not coupled or very weakly coupled to the dark states. 
However, at higher energies the bright state is coupled to the dark states
by various perturbations ${V_{j}}$ which include anharmonic resonances,
coriolis and centrifugal couplings.  These off-diagonal couplings are
responsible for the flow of energy from the initial bright state into the
various dark states. The system is more appropriately described by the
Hamiltonian $H_{0} + \sum_{j}\tau_{j} V_{j}$ in such energy regimes with
the $\tau_{j}$ denoting the strengths of the perturbations $V_{j}$.  Thus,
the experimental signature of IVR in the frequency domain is revealed as a
fragmentation  of a single rovibrational transition into a multiplet. 
Ample evidence exists in the literature indicating that the bright state
is not coupled democratically to all of the possible dark states. Instead
there is a small subset of the  dark states, termed as belonging to the
first tier, that strongly couple to the bright state. This first tier of
states controls the initial stages of the energy flow and hence corresponds
to the shortest timescale in the system dynamics.  They are coupled to
another set of dark states forming the second tier and giving rise to a
second, longer timescale for energy flow.  This pattern of the couplings
leads to a hierarchical tier-model description of IVR in isolated
molecules. For instance, Callegari {\it et. al.} have recently 
observed\cite{benzene} seven different timescales ranging from 100 fs to
2 ns for IVR from the first CH strech overtone of the benzene molecule.

One of the key issues in IVR studies is the nature of the states that make
up the various tiers.  Identifying the nature of the dark states is an
important first step towards the ultimate goal of determining the tier by
tier classification of the states, {\it i.e.} deciding the potential IVR
pathways from the bright state of interest. Theoretically, in order to 
know the destination of the initial excitation, it is necessary to
identify the terms in the molecular Hamiltonian which are responsible for
the coupling between the bright and the dark states.  A molecular
Hamiltonian is needed, but it is well known that determining the {\it
ab initio} potential surface to sufficient accuracy is not routine except
for small molecules.  In order to make progress, a slightly different
viewpoint has been adopted in the literature.  Starting from the
experimentally measured levels spanning the energy range of interest,
an effective spectroscopic Hamiltonian is determined which reproduces the
levels to a certain accuracy.  The anticipation is that such an effective
Hamiltonian can be used to study the eigenstates and dynamics of the
system in the given energy range.  Effective spectroscopic Hamiltonians
have been determined\cite{hcp1,hocl1,ethy,water,qu1,qu2} for a number of
molecules like  phosphaethyne\cite{hcp1} (HCP), hypocholorous
acid\cite{hocl1} (HOCl),  acetylene\cite{ethy}, and water\cite{water}  to
quote a few.  The effective Hamiltonians typically involve anharmonic
resonances which couple the zeroth-order modes of $H_{0}$.  Fairly
detailed quantum, classical, and semiclassical studies using the effective
Hamiltonians have
provided\cite{hcp1,hocl1,ethy1,water1,qs0,qs1,qs2,qs3,qs4,qs5,qs6,qs7,qs8} 
considerable insights into the nature of the eigenstates and IVR.  For
example, the genesis of the isomerization states\cite{hcp1} in HCP and
determining\cite{ethy1} the counter-rotation modes involving the
Hydrogens in C$_{2}$H$_{2}$ have all been possible due to such detailed
quantum-classical correspondence studies.  However, most of these detailed
studies are restricted to two dimensional or quasi-two degree-of-freedom
systems with little\cite{3dwork} or no work on three or higher
degree of freedom systems.  The reasons for this are many fold and mainly
have to do with the fact that correlating quantum eigenstates with
classical Poincar\'e surface of sections, periodic orbits, and
Husimi\cite{husimi} distribution functions from a phase space
perspective is considerably harder for three or more degrees.  From an
eigenstate perspective, it is clear that identifying relevant couplings in
a particular energy range is intimately linked to the possible dynamical
assignments of the highly excited eigenstates.

In principle, the required information is buried in the splitting pattern
as revealed by a eigenstate resolved spectrum.  However, in practice the
task of analyzing the splittings in order to determine even the nature of
the first tier states is nontrivial.  In the happy instances when the
bright state is perturbed strongly by a single state, there is a two-state
scenario, and theoretical as well as experimental techniques can be
brought to bear on the problem.  Examples of such two state interactions
can be found in the experiments by Boyarkin {\it et. al.} on the IVR from
O-H stretch overtones in methanol\cite{meth} and Mork {\it et. al.} on
vibrational mode-coupling in 1,2-difluoroethane\cite{mork}.  Various
level selection schemes have been  proposed\cite{tier1,tier2,tier3,tier4}
which allow one to determine the dark states tier structure for a
specific bright state of interest.  However, as soon as the two state
picture gets superseded by a situation where two or more states are
coupled strongly to the bright state the analysis is quite involved.  A
previous theoretical study\cite{watercpl}  on the assignment of the
highly excited vibrational states of H$_{2}$O provides a good example  of
the complications that can potentially arise even in a three-state
situation.  The complexity of the analysis escalates with increasing
density of states and with the presence of nonperturbative quantum effects
like dynamical\cite{dyntunn} and chaos-assisted tunneling\cite{chaotunn}. 
It is important to note that recent high resolution experiments using
double resonance methods\cite{irir1,irir2,eros,eros2} have made it
possible to elucidate the nature of the dark states that are coupled to
the zeroth-order bright state.  Analyzing the double resonance spectra
with the technique of the hierarchical tree analysis\cite{hier1,hier2},
introduced by Davis, and the traditional perturbation methods yield
important information about the nature of the dark states and the number
of tiers participating in the IVR process.  In addition, with considerable
effort, it is possible to identify the nature of the coupling {\it i.e.,}
anharmonic or coriolis, between the bright state and the first tier
states\cite{propyne}. Such an approach leads to a certain amount of
insight on the important energy flow pathways from a bright state of
interest. Nevertheless, identifying a dominant anharmonic or coriolis
perturbation among the various possibilities is still an important and
difficult problem. 

One of the principal aims of this work is to develop a systematic and
experimentally accessible method to decide the relative importance of
perturbations in a given energy range.  It is also desirable to be able to
analyze multidimensional effective spectroscopic Hamiltonians in terms of
the classical-quantum correspondence without explicit determination of
the periodic orbits.  In particular, this work begins to provide a
theoretical foundation for the central feature that has emerged from the
various experimental and theoretical studies of IVR, {\it i.e.} the
strong correlation between the eigenvalue and eigenfunction fluctuation
measures.  Manifestations of such correlations are evident in the
hierarchical tier structure, power law decay of the survival probability
at intermediate times\cite{rev5,woly,sri,grueb1}, nonergodic energy flow
even in large molecules like Benzene\cite{benzene}, and localized
eigenstates at fairly high energies\cite{water1,ethy1,grueb2}.  The
hierarchical tree analysis\cite{hier1} was one of the first attempts to
take into account this correlation in a systematic manner.  More
recently, Gruebele has addressed the problem by introducing\cite{mfd1}
the matrix fluctuation-dissipation (MFD) method.  Apart from having
numerical advantages, the MFD theorem explicitly highlights the
correlation between spectral intensities and eigenvalues.  Perhaps more
relevant in the context of the present work is the important role played
by the parametric variation of the eigenvalues with changing bright-dark
coupling strength\cite{mfd2}.  Interestingly, many years ago Weissman and
Jortner discovered\cite{weijort} a remarkable correlation between
the topological features of the quantum Poincar\'{e} maps (basically the
Husimi distribution function) and the sensitivity of the energy levels to
the strength of the nonlinear coupling.  In what follows, it will become
clear that the current work provides a quantitative and qualitative basis
for the observations by Weissman and Jortner.

Independently, considerable efforts have been made by the quantum
chaos community towards understanding the origins and mechanisms of
eigenstate localization in fully chaotic
systems\cite{radons,billiard,bsep,cantor,diffus,scar,bog,kaplan}.  In this
extreme limit, quantum ergodicity is conjectured to generate results
consistent with random matrix theories (RMT)\cite{bgs,stechel}.  A key
feature of RMT is the complete absence of correlations between eigenvalue
and eigenfunction properties.  Recently one of us (ST)\cite{stprl}
introduced a sensitive measure of eigenstate localization which is based
on correlations between intensities and eigenlevel velocities. 
Eigenlevel velocities are more properly defined as the slopes of the level
curves generated by changes in some system parameter. It was demonstrated
in  a series of papers\cite{stprl,cllt,lct} that this intensity-level
velocity correlation function provided an ideal measure to explore the
systematic deviations from the predictions of RMT\cite{rmt} due to
localization.  For a detailed introduction to the localization aspects
and as to how it relates to the deviations from RMT predictions we refer
the reader to Ref.~\onlinecite{cllt}.  An attractive feature of a measure based
on the level velocities stems from recent works demonstrating the
fingerprints of a nonlinear resonance in the level dynamics\cite{finger}
and the ability of the level velocities to distinguish between different
regions of the phase space\cite{parisri}.  In a more general context, the
current and recent works\cite{cllt} provide a natural framework to
analyze the spectroscopy of molecules in external fields.  Recent examples
of such ``parametric spectroscopy" experiments include the molecular Stark
effect study\cite{eros,eros2} on 2-propyn-1-ol and the Laser Induced
Fluorescence (LIF) study on NO$_{2}$ in an external magnetic
field\cite{nygard}.  These observations provide us with a strong
motivation to explore the potential usefulness of the correlation measure
in the study of IVR.

The paper is organized as follows.  A brief introduction to the
intensity-level velocity correlation function is provided in section II. 
The main results for the case when the underlying classical dynamics is
fully chaotic, and the RMT expectations are briefly discussed.
In section III, the semiclassical theory for the correlation function is
constructed for classically integrable systems.  Understanding the
integrable case is a necessary component of understanding realistic
spectroscopic Hamiltonians that often possess dynamics more closely
related to the integrable, near-integrable, and/or mixed phase space
regimes.  The near-integrable regime is characterized by the introduction
of resonances, and the applicability of first order classical perturbation
theory.  The mixed phase space regime is characterized by the
co-existence of near-integrable motion and significant chaotic motion
on the same energy surface; {\it i.e.} different initial conditions lead
to different dynamical behaviors.  Explicit expressions are given that
can be verified in great detail, at least in some cases.  The $\hbar$
scalings of the various quantities are deduced as a function of numbers
of degrees of freedom, and shown to be different from the chaotic case. 
In section IV, the intensity-level velocity correlation is computed for
single resonance (classically integrable) and multiresonant (classically
nonintegrable) two degree-of-freedom Hamiltonians.  Section V provides a
brief summary and concludes. 

\section{Intensity-level velocity correlation function : Background}

Consider the effective spectroscopic Hamiltonian $\widehat{H}({\bf a},
{\bf a}^{\dagger};\tau) = \widehat{H}_{0}({\bf n}) + \widehat{H}_{1}
({\bf a},{\bf a}^{\dagger};\tau)$ with $\tau$ representing some parameter
or parameters of interest.  The eigenstates and eigenvalues of
$\widehat{H}$ will be denoted by $|\alpha(\tau)\rangle$ and
$E_{\alpha}(\tau)$ respectively.  In what follows, it is not necessary for
this particular choice of the form of the Hamiltonian.  Thus, $\tau$ could
equally well be a parameter of the zeroth-order Hamiltonian.  The choice
is motivated by a basic interest in determining the localization (or lack
thereof) of eigenstates and hence the perturbations which have a large
effect in a given energy range.  The eigenstates and dynamics associated
with $H_{0}$ are assumed to be well understood which is the generic
situation in IVR studies.  Throughout this article we will focus on
vibrational dynamics and ignore the rotational effects.  Nevertheless, the
basic theory presented below can be applied to the full rovibrational
Hamiltonian.  Using the Heisenberg correspondence rule,\cite{heisen} it is
possible to associate a classical Hamiltonian 
$H({\bf I},{\bm \phi};\tau) = H_{0}({\bf I}) + H_{1}({\bf I},{\bm
\phi};\tau)$ with the quantum $\widehat{H}$.  The form of the
spectroscopic Hamiltonian makes it especially convenient to deal with the
classical Hamiltonian in terms of the action-angle variables corresponding
to the zeroth-order Hamiltonian.  In addition, the classical Hamiltonian
is a nonlinear resonant Hamiltonian representing the various resonant
perturbations which couple the zeroth-order states leading to the flow of
vibrational energy.  Depending on $\tau$ and the energy of interest, the
classical dynamics can exhibit the full dynamical range from integrable to
mixed to chaotic behaviors.

Given a special state $|z\rangle$, the strength
function\cite{scar,stprl,strength} is defined as
\begin{subequations}
\begin{eqnarray}
S_{z}(E,\tau) &=&\frac{1}{2 \pi \hbar} \int_{-\infty}^{\infty} dt
e^{i E t/\hbar} \langle z|e^{-i \widehat{H}(\tau)t/\hbar}|z\rangle \\
&=& \sum_{\alpha} p_{z \alpha}(\tau) \delta(E - E_{\alpha}(\tau))
\end{eqnarray}
\end{subequations}
with $p_{z \alpha}(\tau) = |\langle z|\alpha(\tau)\rangle|^{2}$ being
the overlap intensity.  The strength function had been introduced and
studied earlier by  Heller\cite{strength} in the context of quantum
localization due to eigenstate scarring.  The choice of the special state
is determined by the system of interest.  From a spectroscopic viewpoint,
the special state $|z\rangle$ could be the bright state, and then the
strength function is the observed spectrum.  Alternatively, one can choose
a coherent state corresponding to the phase space image of a certain
bright state of interest.  In such a case, the intensities are essentially
the Husimi function.  In Sect. IV, we will primarily choose the latter
representation in order to highlight the classical-quantum
correspondence of the system.

Beyond scarring, a number of other factors have been found to contribute
towards localization effects.  For instance, transport barriers
like broken separatrices\cite{bsep}, cantori\cite{cantor} or diffusive
motion\cite{diffus} in phase space have been linked to localization.  One
of the hallmarks of localization is the non-democratic response of the
system to parametric variation.  Specifically the `evolution' of the
energy levels of the system, as measured by the level velocities, upon
varying a parameter will not be universal.  Thus, the independence of
eigenvalue and eigenfunction fluctuation measures, a central feature of
RMT, is violated.  An analysis of the correlation between level velocities $\partial
E_{\alpha}(\tau)/\partial \tau$ and the intensities $p_{z \alpha}(\tau)$
leads to a better understanding of the deviations from universal behavior
and gives insights into the nature of any localization.  It is only very
recently\cite{stprl} that such a overlap intensity-level velocity
correlation has been proposed and shown to be a very sensitive measure.
This correlation function is defined as
\begin{equation}
C_{z}(\tau) = \frac{1}{\sigma_{z}\sigma_{E}} \left\langle p_{z\alpha}(\tau)
\frac{\partial E_{\alpha}(\tau)}{\partial \tau}\right\rangle_{E}
\end{equation}
where $\sigma_{z}^{2}$ and $\sigma_{E}^{2}$ are the local variances
of the overlap intensities and level velocities, respectively.  The
brackets denote a local energy average in the neighborhood of $E$.  As
defined, $C_{z}(\tau)$ weights most the states which share common
localization characteristics, and measures the tendency of these levels to
move in a common direction.  Within RMT, it is possible to show\cite{cllt}
that $C_{z}(\tau) = 0 \pm N^{-1/2}$ with $N$ being the effective number
of states in the selected energy range.  It is important to note that the
above result for ergodic systems is true for every choice of $|z\rangle$.
If there is localization, then there is at least one $|z\rangle$ for which
$C_{z}(\tau)$ deviates from zero in a statistically significant way.  The
usefulness of $C_{z}(\tau)$ has been convincingly demonstrated for the
stadium billiard\cite{cllt} and the baker's map\cite{lct}.  A fairly
detailed semiclassical theory for the correlation function when the
underlying classical dynamics is completely chaotic has been
developed\cite{cllt}.  That which is lacking is the theoretical foundation
for integrable, near-integrable, and mixed phase space systems.  

\section{Semiclassical theory for $C_{z}(\tau)$ in the
integrable case}

The prime motivation of introducing and studying $C_{z}(\tau)$
for the chaotic case has to do with uncovering various localization
features of the system and hence deviations from RMT predictions.
Evidently, any localization of eigenstates in the chaotic system is subtle
and somewhat surprising at first sight.  However, localization is the
rule rather than the exception for the integrable to mixed phase
space regimes.  Therefore, it is to be expected to find statistically
significant nonzero values for $C_{z}(\tau)$ in such cases. 

It is demonstrated ahead that the correlation measure extracts useful
information for the non-fully chaotic regimes as well.  The utility stems
from the fact that a perturbation could non-democratically couple certain
states or class of states leading to avoided crossings or like level
movements within the particular class.  Such a preferential coupling is
precisely indicated by $C_{z}(\tau)$ irrespective of the nature of the
underlying classical dynamics.  In context of the present paper, the
effective spectroscopic Hamiltonians typically exhibit a mixed phase
space.  As mentioned previously, the classical limit of the spectroscopic
Hamiltonian is a nonlinear resonant Hamiltonian, and the various
nonlinear resonances manifest themselves in different energy ranges.  It
is quite possible for a molecule in a certain energy range to be
dominated by a single Fermi resonance.  Single resonant classical
Hamiltonians are integrable\cite{licht}, and therefore one has conserved
quantities, called polyads\cite{poly}, associated with the motion.  In a
different energy range, more than one independent resonance can be
important, and classically one has a nonintegrable system.  The entire
range of behavior from integrable to chaotic is therefore possible. 
Analyzing a spectrum with $C_{z}(\tau)$ provides one possible route to
deciphering the important resonances operative in the energy range of
interest.  Moreover, as indicated ahead, there are many other insights,
such as indications of the effective numbers of degrees of freedom found
in the $\hbar$-scalings, that one can gain from a study of the correlation
function which makes it an attractive candidate for exploring IVR in
isolated molecules.

It is important to note that the RMT estimate is only valid for the
extreme fully chaotic case.  In addition, given the richness of
the quantum-classical correspondence for the effective spectroscopic
Hamiltonians, it is both necessary and useful to come up with a
semiclassical theory for $C_{z}(\tau)$ in such regimes.  This section of
the paper closely follows the earlier paper\cite{cllt} concerned with the
correlation function for chaotic systems.  In the current work an
analogus semiclassical theory for the integrable cases is provided.  The
theory for the near-integrable case  will be provided in a later
publication\cite{later}.  

\subsection{Level Velocities}

The purpose of this section is to derive explicit expressions for the
mean and variance of the level velocities.  From these expressions, the
$\hbar$-scaling properties can be deduced.  The analogous expressions and
scalings for the chaotic case have been derived earlier\cite{cllt}.  The
analysis in this section is important  from a classical-quantum
correspondence perspective as well as due to the fact that it can be
translated into a density of states scaling.  To focus on the integrable
case we consider a single $n:m$ resonant Hamiltonian of the form:
\begin{equation}
H({\bf I},{\bm \phi}) = H_{0}({\bf I}) + \tau \sqrt{I_{1}^{m} I_{2}^{n}}
\cos(m\phi_{1} - n\phi_{2})
\end{equation}
It is well known that the above Hamiltonian is integrable because of the
existence of a conserved quantity $I = (n/m)I_{1} + I_{2}$ in addition to
the energy $E$.  This constant of the motion $I$ is called the polyad
number. Since $H({\bf I},{\bm \phi})$ is integrable one can imagine a
canonical transformation $({\bf I},{\bm \phi}) \rightarrow ({\bf K},{\bm
\Phi})$ in terms of which $H = H({\bf K})$.  Stated differently, the
variables $({\bf K},{\bm \Phi})$ are the action-angle varibales for the
single resonance Hamiltonian.  Note that the explicit determination of the
action-angle variables for an arbitrary single resonance Hamiltonian is
very difficult if not impossible.  Joyeux\cite{joy} has given the explicit
solutions for resonances with $m+n \leq 4$ in terms of elliptic functions.
However, not being able to construct the explicit transformation
does not preclude the existence of the action-angle variables for an
integrable system.  As pointed out by Joyeux, with a convenient choice
of the canonical transformation, it is possible to identify one of the 
actions ${\bf K}$ with the polyad number $I$.  Given $H({\bf K})$, it is
possible to define the frequencies ${\bm \nabla}_{{\bf K}} H({\bf K}) =
{\bm \Omega}({\bf K})$.  The single resonance systems have a slight
complication in the sense that the associated phase space has a
separatrix.  This separatrix, which separates resonant and nonresonant
regions in the phase space, cannot be transformed away by a canonical
transformation.  This is reflected in the fact that different action
variables ${\bf K}$ have to be used in the two regions.  The separatrix
structure will be ignored by assuming that only a local region of the
phase space is being investigated in any given case.  A more rigorous
treatment by taking into account separatrices would involve
near-integrable semiclassical theory\cite{amod,ugt}, and that is left for
another work\cite{later}.

In order to determine an expression for the level velocities in terms of
classical quantities and $\hbar$, it is convenient to start from the
smoothed spectral staircase function\cite{cllt,keating,leboeuf}:
\begin{equation}
N_{\epsilon}(E,\tau) = \sum_{\alpha}\theta_{\epsilon}[E - E_{\alpha}(\tau)]
\end{equation}
where $\epsilon$ is an energy smoothing term which will be taken
to be smaller than the mean level spacing.  Differentiating the staircase
function with respect to the parameter $\tau$ and taking the energy
average yields the identity
\begin{equation}
\left \langle \frac{\partial N_{\epsilon}(E,\tau)}{\partial \tau}
\right \rangle_{E} = \bar{d}(E,\tau) \left \langle \frac{
\partial E_{\alpha}(\tau)}{\partial \tau} \right \rangle_{\alpha}
\end{equation}
where $\bar{d}(E,\tau)$ is the mean level density.  Similarly, assuming a
nondegenerate spectrum, it can be shown that
\begin{equation}
\left \langle \left(\frac{\partial N_{\epsilon}(E,\tau)}{\partial \tau}
\right )^{2} \right \rangle_{E} = 
\frac{\bar{d}(E,\tau)}{2 \pi \epsilon} \left \langle \left(
\frac{
\partial E_{\alpha}(\tau)}{\partial \tau} \right)^{2} \right \rangle_{\alpha}
\end{equation}
It is possible to write down a semiclassical expression for the spectral
staircase function as a sum of an average (Weyl) term and an oscillating
term
\begin{equation}
N_{\epsilon}(E,\tau) = \bar{N}(E,\tau) + \widetilde{N}_{\epsilon}(E,\tau)
\end{equation}
For an integrable system, the oscillating part according to the theory of
Berry and Tabor\cite{bt} can be written in terms of the rational periodic
orbits ${\bf M}$ of $H({\bf K})$ as:
\begin{eqnarray}
\widetilde{N}_{\epsilon}(E,\tau) &\approx& \frac{2}{\hbar^{(d - 1)/2}} 
\sum_{\bf M}
A_{{\bf M}}(E,\tau) 
\sin [\sigma_{\bf M}] \nonumber \\
&& \times \exp \left\{\frac{-\epsilon T_{{\bf M}}(E,\tau)}{\hbar} \right\}
\end{eqnarray}
It has been assumed that the curvature of the energy contour is
approximately constant locally and in the sum the ${\bf M} = 0$ term is
excluded. The rational periodic orbits correspond to tori with rotation
numbers $\alpha_{j,k} = \Omega_{j}/\Omega_{k} = \mu_{j}/\mu_{k}$ with
$\mu_{j},\mu_{k}$ being coprime integers.  ${\bm \mu}$ specifies the
primitive periodic orbit and ${\bf M} = r {\bm \mu}$ denotes a specific
topology of the closed orbits with $r$ being the repetition
number\cite{bt}. In the above expression for $\widetilde{N}_{\epsilon}$,
the amplitude and phase factors are given by
\begin{subequations}
\begin{eqnarray}
A_{\bf M}(E,\tau) &=& \frac{1}{2 \pi |{\bf M}|^{(d + 1)/2} 
|\kappa_{\bf M}|^{1/2}}\\
\sigma_{\bf M} &=& \frac{S_{\bf M}(E,\tau)}{\hbar}
-\frac{\pi}{2} \eta_{\bf M} + \frac{\pi}{4} \beta_{\bf M}
\end{eqnarray}
\end{subequations}
where $\kappa_{\bf M}$ is the scalar curvature of the energy contour
$H({\bf K}) = E$. The action corresponding to an orbit of topology ${\bf
M}$ is $S_{{\bf M}} = 2 \pi {\bf K}_{\bf M} \cdot {\bf M}$, the phase
$\eta_{\bf M} = {\bf M} \cdot {\bm \eta}$ with ${\bm \eta}$ being the
Maslov indicies and $\beta_{\bf M}$ is equal to the sign of the 
determinant of the curvature matrix.  The period of the corresponding
orbit is denoted by $T_{\bf M} = r T_{\bm \mu}$. 

The oscillatory contribution to the velocity average can be written
down and evaluated in the $\hbar \rightarrow 0$ limit as
\begin{eqnarray}
\left \langle \frac{\partial \widetilde{N}_{\epsilon}}{\partial \tau}
\right \rangle_{E} &=& 
\frac{2}{\hbar^{(d + 1)/2}} \sum_{\bf M} A_{\bf M}
\left \langle \left(\frac{\partial S_{\bf M}}{\partial \tau} \right)
\cos [\sigma_{\bf M}] \right. \nonumber \\
&& \left. \times \exp \left\{\frac{-\epsilon T_{\bf M}}{\hbar} \right\}
\right \rangle_{E} \nonumber\\
&\approx& \frac{2}{\hbar^{(d + 1)/2}} \sum_{\bf M} A_{\bf M}
\left \langle \left(\frac{\partial S_{\bf M}}{\partial \tau} \right) \right.
\nonumber \\
&& \left. \times \exp \left\{\frac{-\epsilon T_{{\bf M}}}{\hbar} \right\} 
\right \rangle_{E}
\left \langle 
\cos [\sigma_{\bf M}]
\right \rangle_{E}\\
&\approx& 0 \nonumber
\end{eqnarray}
The above result follows from the argument that for a given orbit the
actions and their parametric derivatives are smooth functions of energy.
In the semiclassical limit, the derivatives are uncorrelated from the
phase term which averages to zero in that limit.  Another way to see this
is to observe that in performing the energy averaging by stationary phase
the dominant contribution comes from orbits satisfying
$\sigma^{'}_{\bf M}(E) = 0$ which correspond to the zero period orbits.
By construction, the oscillatory part of the staircase function does not
contain the zero-length orbits and hence the average is vanishingly
small.

Using the expression for $\widetilde{N}_{\epsilon}$ and assuming the
different orbits to be uncorrelated from each other in the semiclassical
limit, one obtains
\begin{eqnarray}
\left \langle \left(\frac{\partial E_{\alpha}(\tau)}{\partial \tau} \right)^{2}
\right \rangle_{E} &=& \frac{2 \pi \epsilon}{\bar{d}}
\left \langle \left(\frac{\partial \widetilde{N}_{\epsilon}}{\partial \tau}
\right)^{2} \right \rangle_{E} \nonumber \\
&=& \frac{4 \pi \epsilon}{\hbar^{(d + 1)} \bar{d}}
\left \langle \sum_{\bf M} A_{\bf M}^{2} \left(\frac{\partial S_{\bf M}}
{\partial \tau} \right)^{2} \right. \nonumber \\
&& \left. \times \exp \left\{\frac{-2 \epsilon T_{\bf M}}
{\hbar} \right\} \right \rangle_{E}
\end{eqnarray}
where a diagonal approximation has been used for the sum over rational
tori. The off-diagonal terms will contribute to the variance but the
dominant contributions and $\hbar$-dependence in the semiclassical limit
is expected to derive from the diagonal term.  For integrable systems,
it is reasonable to assume\cite{note} the relation
\begin{equation}
\left \langle \left(\frac{\partial S_{\bf M}}
{\partial \tau} \right)^{2} \right \rangle_{\bf M} \approx \zeta(E,\tau) T^{2}
\end{equation}
The distinction between $T^2$ and $T$ (relevant to the chaotic case) in
the above expression is analogous to the distinction between ballistic
and diffusive behaviors, respectively.  In analogy to the chaotic case, we
invoke the appropriate Hannay-Ozorio sum rule\cite{hozsum}
\begin{equation}
\sum_{\bf M} A_{\bf M}^{2} \rightarrow \frac{\hbar^{d} \bar{d}}{2 \pi}
\int \frac{dT}{T^{2}}
\end{equation}
to perform the sum over the rational tori.  As a result we obtain the
dimensionless variance $\tilde{\sigma}^{2}_{E}$ as
\begin{equation}
\tilde{\sigma}^{2}_{E} = \bar{d}^{2} 
\left \langle \left(\frac{\partial E_{\alpha}(\tau)}{\partial \tau} \right)^{2}
\right \rangle_{E}
= \zeta(E,\tau) \bar{d}^{2} \sim \hbar^{-2d}
\end{equation}
The relation between the mean square action changes and the mean square
level velocities is straightforward involving only the density of states
to complete the connection.  The $\hbar$ dependence drops out easily since
the mean level density $\bar{d} \sim \hbar^{-d}$ for a
$d$-degree-of-freedom  system.  This is to be compared to the scaling
$\tilde{\sigma}^{2}_{E} \sim \hbar^{-(d + 1)}$ previously established for
chaotic systems\cite{cllt}.  Note that the different $\hbar$ scalings are
consistent with the notion that chaotic systems' spectra fluctuate less
and are more rigid (constrained) than their integrable systems'
counterparts.  Furthermore, both the effective numbers of degrees of
freedom and the nature of the dynamics is reflected in the scaling laws. 
This is a general feature arising over and over.  As a consequence of the
above result for integrable systems, the variance of the level velocities
scales as $\sigma^{2}_{E} \sim O(\hbar^{0})$, and is equal to the
variance of the action changes averaged over the appropriate tori. 
Numerical results in the next section, obtained from application to
integrable single resonance Hamiltonians, confirm the derived
$\hbar$-scaling for $\tilde{\sigma}^{2}_{E}$.  The full expression has
been verified for the rectangular billiard, but only the scalings will be
presented in this paper.  

\subsection{Overlap intensities}

In this section a semiclassical expression for the strength function is
derived. The derivation is firmly rooted in the action-angle space and
consistent with the Berry-Tabor rational periodic orbit formalism. The
$\hbar$ scaling  of the intensity is derived and shown to be different
from that of its chaotic counterpart. 

As with the level velocities it is possible to decompose the strength
function $S_{z}(E,\tau)$ into an average and an oscillatory part
\begin{equation}
S_{z}(E,\tau) =
\bar{S}_{z}(E,\tau) + \widetilde{S}_{z}(E,\tau)
\end{equation}
where $| z \rangle$ is a coherent state and $p_{z \alpha} = 
|\langle z|\alpha(\tau)\rangle|^{2}$.  The average or the smooth part is
the contribution from zero length trajectories.  It can also be viewed as
the Fourier transform of the extremely rapid initial decay due to 
the shortest time scale of the dynamics\cite{strength}.

The average intensities can be written in terms of the smooth part of the
strength functions as:
\begin{equation}
\langle p_{z \alpha}(\tau) \rangle_{\alpha} = \frac{1}{\bar{d}}
\bar{S}_{z}(E,\tau)
\end{equation}
The quantity of main interest is thus 
\begin{equation}
\bar{S}_{z}(E,\tau) = \frac{1}{(2 \pi \hbar)^d} 
\int A_{w}({\bf K},{\bm \Phi})
\delta [E - H({\bf K})] d{\bf K} d{\bm \Phi}
\end{equation}
$A_{w}({\bf K},{\bm \Phi})$ is the Wigner transform of 
a Gaussian wavepacket centered at $({\bar{\bf K}},{\bar{\bm \Phi}})$
and can be represented locally in the action angle space $({\bf K},{\bm \Phi})$
by
\begin{equation}
A_{w}({\bf K},{\bm \Phi}) = 2^{d} 
\exp \left[
-(\Delta {\bm \Phi})^{2}
-\frac{1}{\hbar^{2}} (\Delta {\bf K})^{2} \right]
\end{equation}
The vectors $\Delta {\bm \Phi}$ and $\Delta {\bf K}$ have components
$(\Delta {\bm \Phi})_{j} = (\Phi_{j} - \bar{\Phi}_{j})/\sigma_{j}$
and $(\Delta {\bf K})_{j} = \sigma_{j}(K_{j} - \bar{K}_{j})$
respectively where $\sigma_{j}^{2} = \hbar/2\bar{K}_{j}$.  In the above
expression, a dimensionless variable has been scaled to one in order to
have equal uncertainty scalings in the angle and action variables.  As
the Hamiltonian is independent of the angles, the angle integrals can be
performed in the semiclassical limit.  A change of variables
$\hbar {\bf z} = \Delta {\bf K}$ leads to the expression
\begin{equation}
\bar{S}_{z} \approx \frac{1}{\pi^{d/2}} \int d{\bf z} e^{-{\bf z}^{2}}
\delta \left[E - H\left(\hbar {\bf z} \cdot {\bm \sigma}^{-1} 
+ \bar{\bf K}\right)\right] 
\end{equation}
whose evaluation gives:
\begin{eqnarray}
\bar{S}_{z}
&\approx& \frac{1}{\pi^{d/2}} \int d{\bf z} e^{-{\bf z}^{2}} \delta
\left[  \hbar^{1/2} \sum_j  \sqrt{2 \bar{K}_{j} \Omega_{j}^{2}} z_{j}
\right] \nonumber \\
&=&  \left( 2 \pi \hbar \sum_{j=1}^d \bar{K}_{j} \Omega_{j}^{2}
\right)^{-1/2}
\end{eqnarray}
The final form results from expanding the Hamiltonian about $\bar{\bf K}$
and evaluating the integral at $E = H(\bar{\bf K})$, which thus pertains
to the peak of the strength function.  The $\hbar$-scaling can be
extracted giving
\begin{equation}
\langle p_{z \alpha}(\tau) \rangle \sim \hbar^{d - 1/2}
\end{equation}
It is important that the derived $\hbar$-scaling for the intensities is
independent of the nature of the dynamics, {\it i.e.} integrable or
chaotic, as must happen. 

The oscillatory part of the strength function is a dynamical object and
can be expressed in the action-angle space as
\begin{equation}
\widetilde{S}_{z}(E,\tau) = \frac{-1}{\pi} {\rm Im} \int d{\bm \Phi}
d{\bm \Phi}' \langle z | {\bm \Phi} \rangle G({\bm \Phi},{\bm \Phi}';E)
\langle {\bm \Phi}'|z \rangle
\end{equation}
where $G({\bm \Phi},{\bm \Phi}';E)$ is the energy Green's function.
The semiclassical expression for the Green's function is\cite{amod}
\begin{eqnarray}
G({\bm \Phi},{\bm \Phi}';E) &\approx& 
\frac{1}{i \hbar (2 \pi i \hbar)^{(d - 1)/2}}
\sum_{{\bf K}} |D_{\bf K}|^{1/2} \nonumber \\
&& \times \exp \left(\frac{i}{\hbar} S_{\bf K}
({\bm \Phi},{\bm \Phi}';E) - \frac{i \pi \nu_{\bf K}}{2} \right)
\end{eqnarray}
where the sum is over all possible trajectories (associated with tori ${\bf K}$)
that go from ${\bm \Phi}'$ to ${\bm \Phi}$ at an energy $E$ and their
repetitions.  $S_{\bf K}({\bm \Phi},{\bm \Phi}';E)$ is the action
associated with a trajectory (including the repetitions) and $D_{\bf K}$
is the usual stability determinant of order $(2d + 1) {\times} (2d + 1)$
given by
\begin{equation}
 D_{\bf K} = \left| \begin{array}{cc}
    \frac{\partial^{2} S_{\bf K}}{\partial {\bm \Phi} \partial {\bm \Phi}'} &
    \frac{\partial^{2} S_{\bf K}}{\partial {\bm \Phi} \partial E} \\
    \frac{\partial^{2} S_{\bf K}}{\partial E \partial {\bm \Phi}'} &
    \frac{\partial^{2} S_{\bf K}}{\partial E^{2}}
   \end{array}\right| 
\end{equation}
The product of the wavefunctions in the coherent state basis can be
obtained via an inverse Wigner transform\cite{amod} of the phase space
Gaussian.  The inverse transform is evaluated by stationary phase
to obtain
\begin{eqnarray}
\langle z|{\bm \Phi}\rangle \langle {\bm \Phi}'|z\rangle &\approx&
\frac{1}{\pi^{d/2} \prod_{j} \sigma_{j}}
\exp \left[-\frac{1}{2} (\Delta {\bm \Phi})^{2} - \frac{1}{2}
(\Delta {\bm \Phi}')^{2} \right. \nonumber \\
&+& \left. \frac{i}{\hbar} \bar{\bf K} \cdot
({\bm \Phi} - {\bm \Phi}') \right]
\end{eqnarray}
The evaluation by stationary phase is valid if the location of the
Gaussian is away from the edges of the phase space. This is consistent
with the fact that we have used a local form of the coherent state in
the action-angle coordinates.

The Gaussian localization in action-angle space implies that the 
dominant contribution to $\widetilde{S}_{z}$ arises from trajectories
which are closed on the torus ${\bf K}$ {\it i.e.,} ${\bm \Phi} \approx
{\bm \Phi}' = \bar{\bm \Phi}$.  In other words, the dominant contribution
comes from rational tori ${\bf K}_{\bf M}$.  This motivates an
expansion of the action as
\begin{equation}
S_{\bf K}({\bm \Phi},{\bm \Phi}';E) \approx 
S_{\bf M}(\bar{\bm \Phi},\bar{\bm \Phi};E)
+ {\bf K} \cdot ({\bm \Phi} - {\bm \Phi}')
\end{equation}
In the above expansion the quadratic terms have been neglected.  As
discussed earlier, the rational torus action $S_{\bf M} =  2 \pi {\bf
K}_{\bf M} \cdot {\bf M}$.  This is reasonable for an integrable system
since the initial Gaussian in phase space simply shears with time.  Using
the above expansion of the action, the integrals over the angles can
be performed by stationary phase giving
\begin{eqnarray}
\widetilde{S}_{z}(E,\tau) &\approx& \frac{2^{d} \prod_{j} \sigma_{j}}
{\hbar (2 \pi \hbar)^{(d - 1) / 2}} \sum_{\bf M} |D_{\bf M}|^{1/2}
\cos[\sigma_{\bf M}] \nonumber \\
&& \times e^{-(\Delta {\bf K}_{\bf M})^{2}/\hbar^{2}}
\end{eqnarray}
The above result for the oscillatory part of the strength function
can be used to obtain the variance of intensities as
\begin{subequations}
\begin{eqnarray}
\sigma^{2}_{z} &=& \frac{2 \pi \epsilon}{\bar{d}} 
\langle {\widetilde S}^{2}_{z}(E,\tau) \rangle_{E} \\
&=& \frac{\epsilon 2^{2d - 1} \prod_{j} \sigma_{j}^{2}}
{\bar{d} (2 \pi)^{d - 2} \hbar^{d + 1}}
\left \langle \sum_{\bf M} |D_{\bf M}| e^{-2 (\Delta {\bf K}_{\bf M})^{2}/
\hbar^{2}} \right. \nonumber \\
&& \left. \times e^{-2 \epsilon T_{\bf M}/\hbar} \right \rangle_{E}
\end{eqnarray}
\end{subequations}
where we have used the diagonal approximation for the sum over ${\bf M}$.
Since the trace of the Green function is the density of states given by
Berry and Tabor\cite{bt} and the density of states is the energy
derivative of the spectral staircase function, then $|D_{\bf M}| = T_{\bf
M}^{2} A_{\bf M}^{2}/ (2 \pi)^{d-1}$.  Hence, given the Hannay-Ozorio sum
rule\cite{hozsum} for $\sum_{\bf M} A_{\bf M}^{2}$ stated in the
previous section, here
\begin{equation}
\sum_{\bf M} |D_{\bf M}| \rightarrow \frac{\hbar^{d} \bar{d}}{(2 \pi)^d
} \int dT
\end{equation}
It follows that
\begin{equation}
\sigma_{z}^{2} \approx \frac{\prod_{j} \sigma_{j}^2}{\pi^{2d - 2}} 
\left \langle e^{-2 (\Delta {\bf K}_{\bf M})^{2}/\hbar^{2}}
\right \rangle_{\bf M}
\end{equation}
which again has been verified for the rectangular billiard, but will not be
shown here.  The leading order $\hbar$-scaling of the intensity variance is 
\begin{equation}
\sigma_{z}^{2} \sim \hbar^{d}
\end{equation}
In the chaotic case\cite{error}, the RMT expectation is $\sigma_{z}^{2}
\sim \hbar^{2d - 1}$ which follows from the scaling of $\langle p_{z
\alpha} \rangle$.  The intensities follow a Porter-Thomas distribution,
which has only one parameter given by its mean; the variance is always
twice the square of the mean.  Again, the integrable limit scaling
of the variance of the intensities is different from the chaotic limit
which is a manifestation of the differing nature of the underlying
classical dynamics.  It follows that because the variance is on a much
greater scale, the distribution of strengths is necessarily singular for
the integrable case in the limit of $\hbar\rightarrow 0$ ({\it i.e.}
mostly near zero values except for a few extremely large values).
 
\subsection{Correlation function}

The semiclassical theory for overlap intensities and level velocities
outlined in the previous subsections can now be used to construct the
correlation function.  Beginning with the covariance, it can be shown
that 
\begin{eqnarray}
Cov_{z}(\tau) &=&
\left \langle p_{z \alpha}(\tau) \frac{\partial E_{\alpha}(\tau)}{
\partial \tau} \right \rangle_{\alpha} \nonumber \\ 
&=&\frac{2 \pi \epsilon}{\bar{d}} \left \langle S_{z}(E,\tau) \frac{
\partial N(E,\tau)}{\partial \tau} \right \rangle_{E}
\end{eqnarray}
The semiclassical expressions for the strength function and the
parametric derivative of the staircase function give
\begin{eqnarray}
Cov_{z}(\tau) &=&
\frac{\epsilon 2^{d} \prod_j \sigma_j}{\bar{d} \hbar^{d + 1} 
(2 \pi)^{(d - 3) / 2}}
\left \langle \sum_{\bf M}
|A_{\bf M}| |D_{\bf M}|^{1/2} \left(\frac{\partial S_{\bf M}}
{\partial \tau}\right) \right. \nonumber \\
&& \left. \times e^{-(\Delta {\bf K}_{\bf M})^{2}/\hbar^{2}}
e^{-2 \epsilon T_{\bf M}/\hbar} \right \rangle_{E}
\end{eqnarray}
Again, the diagonal approximation has been made in the above expression.
Following arguments parallel to those in the previous section leads to
the expression
\begin{eqnarray}
Cov_{z}(\tau) &=&
\frac{2^{d} \prod_j \sigma_{j}}{(2 \pi)^{d - 1} \tau_{H}}
\int dT \frac{1}{T} e^{-2T/\tau_{H}} \left \langle 
\left(\frac{\partial S_{\bf M}}{\partial \tau}\right) \right. \nonumber \\
&& \times \left. e^{-(\Delta {\bf K}_{\bf M})^{2}/\hbar^{2}}
\right \rangle_{\bf M}
\end{eqnarray}
where $\tau_{H} = \hbar/\epsilon$ is the Heisenberg time. 
In order to make further progress in understanding the correlation
function, it is necessary to know the time dependence of the 
${\bf M}$-averaged expression.  It is important to note that the Gaussian
weighting factor will not decouple from the paramteric action derivative
at long times.  Assuming that the parametric derivative $\partial {\bf
K}_{\bf M}/\partial \tau$ to be a weak function of the period gives the
approximation
\begin{equation}
\left \langle\left(\frac{\partial S_{\bf M}}{\partial \tau}\right)
e^{-(\Delta {\bf K}_{\bf M})^{2}/\hbar^{2}}
\right \rangle_{\bf M} \approx T \xi (\tau)
\end{equation}
leading to the final result
\begin{equation}
Cov_{z}(\tau) \approx \frac{\xi (\tau) \prod_{j} \sigma_{j}}{\pi^{d - 1}}
\end{equation}
Combining this with the results of the previous subsections generates the
semiclassical expression for the correlation function.  It is
\begin{equation}
C_{z}(\tau) \approx \frac{\xi (\tau)} {\zeta^{1/2} (E,\tau) \left\langle
e^{-2 (\Delta {\bf K}_{\bf M})^{2}/\hbar^{2}}\right\rangle_{\bf M}^{1/2}}
\end{equation}
Interestingly, there is classical information represented in the
correlation function that is not in the covariance;
therefore considering the separate ``components'' of the correlation
function (or considering both the covariance and correlation functions)
gives the greatest information about the system.  In the semiclassical
limit, the above result implies that $Cov_{z}(\tau) \sim \hbar^{d/2}$ for
integrable systems whereas the correlation $C_{z}(\tau) \sim \hbar^{0}$
to leading order.  This last result had to emerge from the semiclassical
calculations.  Otherwise strong localization could not persist in
integrable systems in the limit as $\hbar \rightarrow 0$, and
multidimensional WKB theory could not work.

\section{Application to a model effective Hamiltonian}

In order to illustrate the usefulness of the intensity-level velocity
correlation measure introduced in the previous sections, consider 
a simple two dimensional effective Hamiltonian
\begin{equation}
\widehat{H}({\bf n},{\bf a},{\bf a}^{\dagger}) = \widehat{H}_{0}({\bf n}) + 
\tau_{11} \widehat{H}_{1:1}({\bf a},{\bf a}^{\dagger})
+\tau_{21} \widehat{H}_{2:1}({\bf a},{\bf a}^{\dagger})
\end{equation}
where $\widehat{H}_{0}({\bf n})$ is the zeroth-order, anharmonic,
vibrational Hamiltonian given by
\begin{equation}
\widehat{H}_{0}({\bf n}) = \omega_{1}N_{1} + \omega_{2} N_{2} +
x_{11} N_{1}^{2} + x_{22} N_{2}^{2}
\end{equation}
The zeroth-order frequencies of the two modes are denoted by
$\omega_{1},\omega_{2}$.  The mode occupation numbers are $N_{j} \equiv
(n_{j} + 1/2)$  and the mode anharmonicities are denoted by $x_{jj}$.
The full Hamiltonian is multiresonant in that the zeroth-order anharmonic
modes are coupled by a $1:1$ and a $2:1$ resonant perturbation 
defined in terms of the operators ${\bf a},{\bf a}^{\dagger}$ as
\begin{eqnarray}
\widehat{H}_{1:1}({\bf a},{\bf a}^{\dagger}) &=& a_{1}^{\dagger} a_{2} +
a_{2}^{\dagger} a_{1} \nonumber \\
\widehat{H}_{2:1}({\bf a},{\bf a}^{\dagger}) &=& a_{1}^{\dagger} a_{2} a_{2} +
a_{2}^{\dagger} a_{2}^{\dagger} a_{1}
\end{eqnarray}
The parameters $\tau_{11}$ and $\tau_{21}$ measure the strength of the
corresponding resonant perturbations.  The above spectroscopic Hamiltonian
$\widehat{H}$  arises in many instances in molecular spectroscopy.
For example, with $\tau_{21} = 0$ and for identical oscillators, the
effective Hamiltonian models\cite{water1} the stretching modes of
H$_{2}$O.  On the other hand with $\tau_{11} = 0$, the effective
Hamiltonian models the Fermi resonant interactions between the stretch
and the bend modes in HCX$_{3}$ molecules\cite{qu1}.  We have chosen a
multiresonant Hamiltonian to model systems where two zeroth-order modes
of interest can get coupled by one or both of the resonances depending on
the pertinent energy range.  The parameter values chosen for this work are
$\omega_{1} = 1.0, \omega_{2} =  0.8, x_{11} = -0.04$, and $x_{22} =
-0.02$.

The classical limit Hamiltonian corresponding to the quantum $\widehat{H}$
can be easily constructed using the Heisenberg correspondence
rule\cite{heisen} and shown to be the following nonlinear, multiresonant
Hamiltonian:
\begin{eqnarray}
H({\bf I}, {\bm \phi}) &=& \omega_{1}I_{1} + \omega_{2} I_{2} +
x_{11} I_{1}^{2} + x_{22} I_{2}^{2} \nonumber \\
&+& 2 \tau_{11} (I_{1} I_{2})^{1/2}
\cos (\phi_{1} - \phi_{2}) \nonumber \\
&+& 2 \tau_{21} (I_{1} I_{2}^{2})^{1/2}
\cos (\phi_{1} - 2 \phi_{2})
\end{eqnarray}
It is well known that with either $\tau_{11}$ or $\tau_{21}$ equal to
zero the classical Hamiltonian is still integrable whereas with both the
coupling strengths nonzero, the system is nonintegrable.  For the
choice of the parameter values made above and noting that the model
Hamiltonian is two dimensional, the density of states is expectedly small.
In order to simulate a high density of states situation we will make use
of the following scaling\cite{qs5} to reduce the effective value of $\hbar
\rightarrow \hbar/c$:
\begin{eqnarray}
\omega_{j} &\rightarrow& \bar{\omega}_{j} = \omega_{j}/c,\ \ 
x_{ij} \rightarrow \bar{x}_{ij} = x_{ij}/c^{2}, \nonumber \\
\tau_{11} &\rightarrow& \bar{\tau}_{11} = \tau_{11}/c, \ \ 
\tau_{21} \rightarrow \bar{\tau}_{21} = \tau_{21}/c^{3/2}
\end{eqnarray}
Note that for most of the results presented in this work we have used
$\hbar=1$.  Elsewhere, the scaling will be explicitly given.

In computing the correlation functions shown below, the level velocities
have been constrained to be zero-centered\cite{cllt}.  This is achieved by
subtracting out the mean level velocity, in the energy range of interest,
from the individual level velocities; this results in no loss of
generality.  Also, the energy window is chosen such that the classical
dynamics are essentially the same throughout the range.  This is essential
since any bifurcation in the system will alter\cite{parisri} the level
velocity structure.  The problem of qualitative changes in the
correlations due to the various bifurcations will not be addressed here.
The ``bright state" $|z\rangle$ is chosen to be a coherent state
centered on $(J,\psi)$ with $J = (I_{1}-I_{2})/2$ and $\psi =
(\phi_{1}-\phi_{2})$.  This choice of $|z\rangle$ is convenient for
studying the system's classical-quantum correspondence via comparisons to
the appropriate classical Poincar\'{e} surface of sections.  The details
for the construction of the surface of sections and determining the
intensities $p_{z\alpha}$ can be found in an earlier work\cite{water1}.
The level velocities are computed by invoking the Hellman-Feynman theorem
and given by
\begin{equation}
\frac{\partial E_{\alpha}}{\partial \tau_{j}} = \langle \alpha |
\widehat{V}_{j} | \alpha \rangle
\end{equation}
Note the important point that the level velocities themselves are
basis independent objects.  The energy range used for computing the
correlation is fixed to be $[8,9]$.  For this choice of the energy range,
the eigenvalues and eigenstates are well converged in the zeroth-order
number basis.  The choice of the energy range is also motivated by
the fact that the two resonances overlap without leading to large
scale stochasticity in the classical phase space.

\subsection{Example of integrable cases}

The multiresonant Hamiltonian has two integrable, single resonance
subsystems that can be obtained by setting $\tau_{11} \neq 0,\tau_{21} = 0$ and
$\tau_{11} = 0, \tau_{21} \neq 0$.  In the former case one obtains a
$1:1$ system and in the latter case a $2:1$ system is obtained.  Studying
these integrable subsystems serves two main purposes. First, the
evaluation of the correlation function $C_{z}(\tau)$ helps provide a
baseline for the later interpretations in the  nonintegrable regime. 
Secondly, the $\hbar$-scalings predicted for $\sigma_{\alpha}$ and
$\sigma_{E}$ can be verified.  In principle, the full semiclassical
expressions could be verified.  However, that would require the
calculation of a large set of periodic orbits which would be time
consuming for this system; the expressions have been verified for the
simpler system of a rectangular billiard.  Note that at the outset it is 
expected that the correlation will be significant for many choices of the 
coherent states $|z\rangle$.  Our concern, however, is to determine the nature
of $C_{z}(\tau)$ for the integrable systems and compare it to the
classical phase space structures.

\begin{figure}
\includegraphics*[width=3in,height=5in]{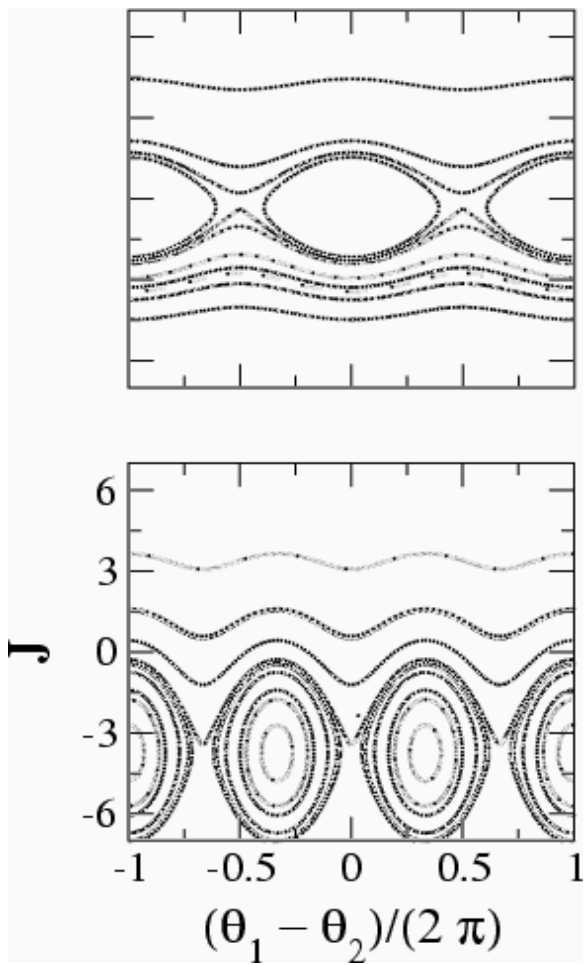}
\caption{Poincar\'{e} surface of sections for the integrable
subsystems at energy $E = 8.5$.
The top panel corresponds to the $1:1$ case and the
bottom panel to the $2:1$ case.}
\label{fig 1}
\end{figure}

\begin{figure}
\includegraphics*[width=3in,height=5in]{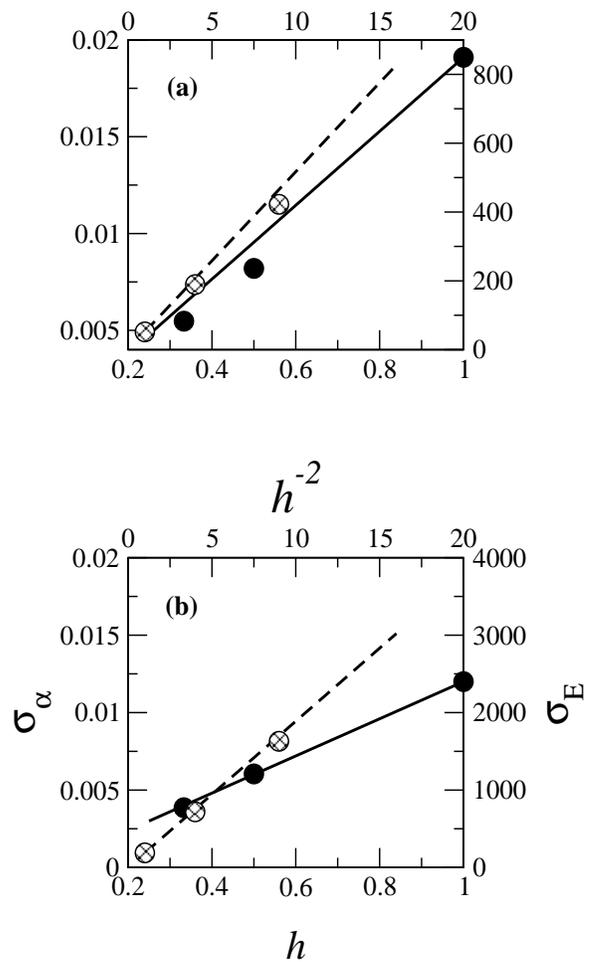}
\caption{$\hbar$-scaling for the intensity variance ($\sigma_{\alpha}$)
shown as solid circles
and the dimensionless velocity variance ($\widetilde{\sigma}_{E}$)
shown as open circles. The lines correspond to the theoretical
estimates.
(a) integrable $1:1$ with $\tau_{11} = 10^{-2}, \tau_{21} = 0$.
(b) integrable $2:1$ with $\tau_{11} = 0, \tau_{21} = 10^{-2}$.}
\label{fig 2}
\end{figure}

Throughout this section the nonzero coupling strength is set equal
to 10$^{-2}$.  In Fig.~(\ref{fig 1}), we show the classical $(J,\psi)$
surface of section at an energy $E = 8.5$ for the integrable $1:1$ and the
$2:1$ cases, respectively.  The appropriate resonant and nonresonant
regions for the two cases can be clearly identified.  In the energy range
of interest, there are no bifurcations in the classical phase space.  In
Fig.~(\ref{fig 2}), we show the intensity and dimensionless level
velocity variances as a function of $\hbar$ and $\hbar^{-2}$,
respectively (so that the points lie on a straight line), for the two
subsystems.  In the $1:1$ case, the coherent state is located  at
$|z_{A}\rangle = (0,-1)$ in the $(J,\psi)$ surface of section whereas in
the $2:1$ case, the coherent state is  located at $|z_{B}\rangle =
(-0.5,-3.75)$.  The particular choices of the coherent states are made
such that they are located inside the resonance zones and close to the
appropriate stable periodic orbits. The agreement with the theoretical
predictions is fairly good and other choices of $|z\rangle$ yield similar
results except near the edges of the $(J,\psi)$ plane.  Although not shown
here, the scaling of the average intensities $\langle p_{z\alpha}
\rangle$ is also consistent with the theoretical values derived in the
previous section.

\begin{figure}

\subfigure[]
{\includegraphics*[width=4.0in,height=2.0in]{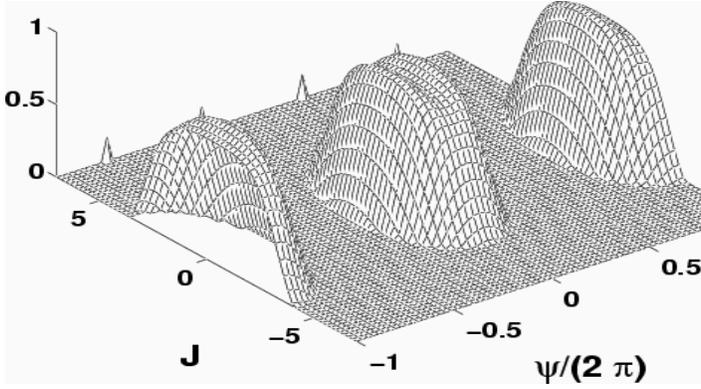}}

\subfigure[]
{\includegraphics*[width=3.5in,height=1.5in]{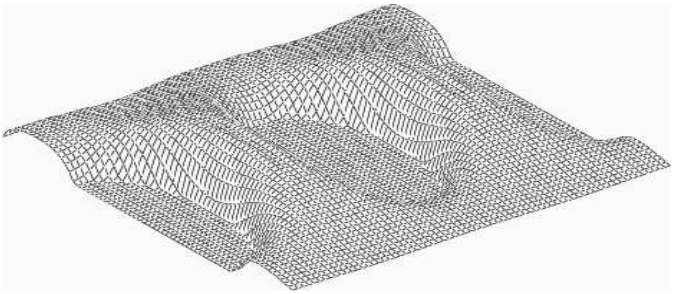}}

\caption{(a) Positive part of the correlation $C_{z}(\tau_{11})$ for
the integrable $1:1$ subsystem with each point representing a
possible coherent state $|z\rangle$ in the $(J,\psi)$ plane.
The axes are $(J,\psi)$ with identical
limits as in the surface of section in figure 1.
The $z$-axis ranges from $0$ to $1$. Note that all the
figures illustrating the correlation function have the same range
for the axes as in this figure.
(b)Negative part of the correlation $C_{z}(\tau_{11})$ for
the integrable $1:1$ subsystem.}
\label{fig 3}
\end{figure}

\begin{figure}

\subfigure[]
{\includegraphics*[width=3.5in,height=1.5in]{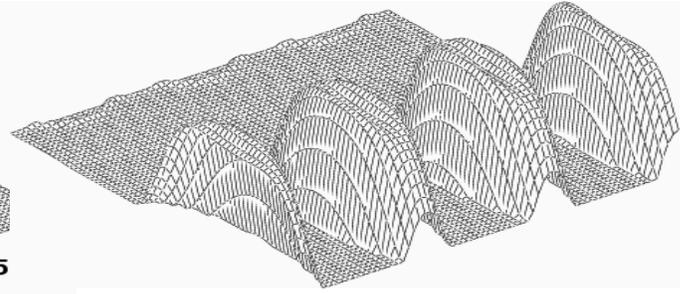}}

\subfigure[]
{\includegraphics*[width=3.5in,height=1.5in]{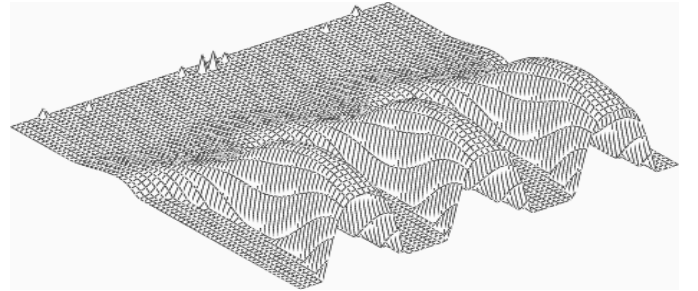}}

\caption{(a) Positive and (b) negative part
of the correlation $C_{z}(\tau_{21})$ for
the integrable $2:1$ subsystem.}
\label{fig 4}
\end{figure}

We now turn our attention to the intensity-level velocity correlations
and compute them as a function of the coherent states spanning the
surface of section; each correlation result is plotted at the phase space
center of the coherent state being used, and which thereby gives a
full phase space portrait of the correlation function's behavior.  In
Fig.~(\ref{fig 3}), we show the positive and negative parts of the
correlation $C_{z}(\tau_{11})$, respectively.  The prime reason for
showing the positive and negative contributions of $C_{z}(\tau_{11})$
separately is that this directly reflects the phase space localization
features of states with positive and negative level velocities.  As
mentioned in the introduction, the relative sign of the level velocities
in integrable and mixed systems is intimately connected to the
localization of the states about stable and unstable structures in the
phase space.  For $\hbar = 1$, the number of states in the energy range is
$27$ for both the $1:1$ and the $2:1$ systems.  Despite the low density
of states, Fig.~(\ref{fig 3}a) shows large positive, statistically
significant values for $C_{z}(\tau_{11})$ around the stable periodic
orbits.  Similarly from Fig.~(\ref{fig 3}b), notice the large values for
the negative correlation around the unstable periodic orbit; the positive
and negative peak values are 
$0.92$ and $0.42$ respectively.
The results for the integrable $2:1$
case are shown in Fig.~(\ref{fig 4}).  Once again the correlation
$C_{z}(\tau_{21})$ takes on large positive ($0.83$) 
and negative ($0.65$) values 
at the stable and unstable periodic orbits respectively.  The
strong positive peaks in the correlations are reminiscent of similar
structures arising due to the bouncing ball states in the stadium
system\cite{cllt}.  Although the stadium is completely chaotic, the
analogy to the present integrable case is appropriate.  This has to do
with the fact that a previous work\cite{finger} has shown that the states
localized in the center of the resonance zone exhibit linear parametric
motion.  It is well known\cite{takhas} that the bouncing ball states also
exhibit linear parametric motion and, indeed, a WKB quantization of such
states leads to\cite{cllt} $\widetilde{\sigma}_{E} \sim \hbar^{-2}$ which
compares favorably with our integrable system theory.  Strong
correlations at the classical fixed points have been observed 
earlier\cite{lct} in the context of the bakers and the standard map.  The
present results are consistent with the previous work, and a significant
observation is that the correlation is a faithful representation of the
relevant resonant portion of the phase space.  In other words, depending
on the parameter being varied, the corresponding region of the classical
phase space is highlighted.  A particularly interesting feature of the
correlations in Figs.~(\ref{fig 3},\ref{fig 4}) has to do with a
noticeable dip at the classical fixed points.  This seems to be a fairly
general feature associated with fixed  points in the classical phase
space.  In a sense, the correlation, constructed with quantum eigenvalues
and eigenstates, is capable of locating the periodic orbits of the
system.  We do not intend to give a detailed account of this feature
except to mention that it arises from the large intensity variance
$\sigma_{\alpha}$ at the fixed points.

\subsection{Example of mixed phase space case}

Consider the full multiresonant Hamiltonian with $\tau_{11} = \tau_{21} =
10^{-2}$.  The theory developed in this work is not strictly applicable
in this regime, although if prior experience can be used as a guide, it
will be relevant to the regular phase space regions or KAM islands (as
will be seen ahead).  On the other hand, a mixed phase space is generic to
many systems, and it is essential to understand the structure of the
correlation function in such cases.  It is important to note that a
theoretical basis for the mixed phase space case is exceedingly
difficult, and currently the next best approach is to attempt\cite{later}
a description for the near-integrable systems.

\begin{figure}
\includegraphics*[width=3.5in,height=3.5in]{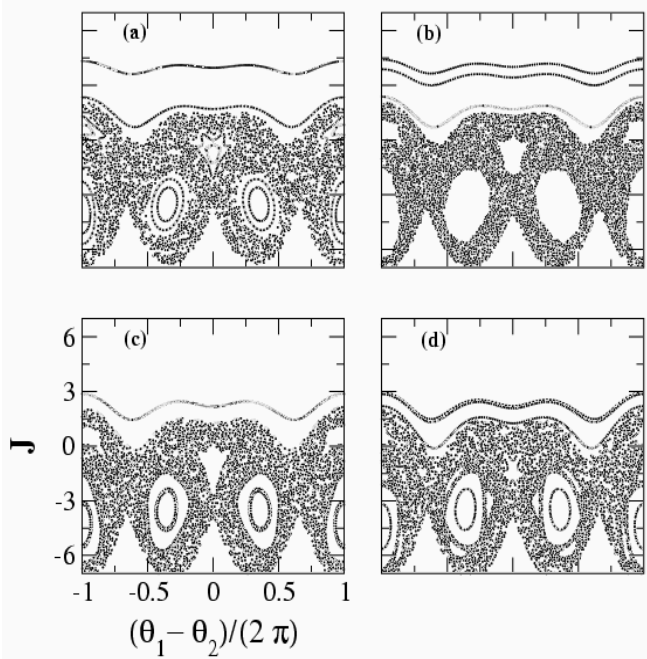}
\caption{Poincar\'{e} surface of sections for the multiresonant Hamiltonian
with $\tau_{11} = \tau_{21} = 10^{-2}$. (a) $E = 8.00$ (b) $E = 8.25$
(c) $E = 8.75$ and (d) $E = 9.00$.}
\label{fig 5}
\end{figure}

\begin{figure}
\subfigure[]
{\includegraphics*[width=3.5in,height=1.5in]{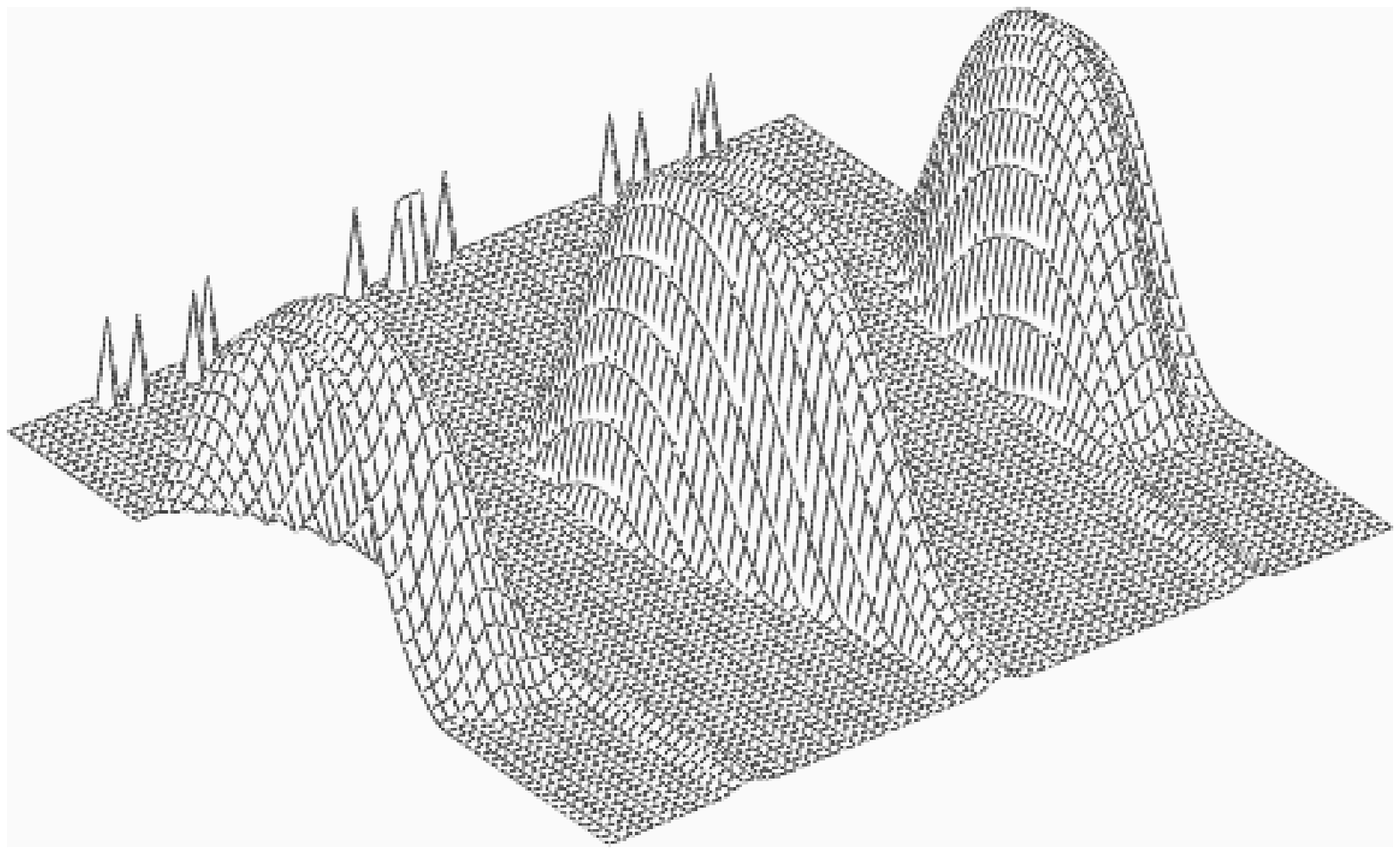}}

\subfigure[]
{\includegraphics*[width=3.5in,height=1.5in]{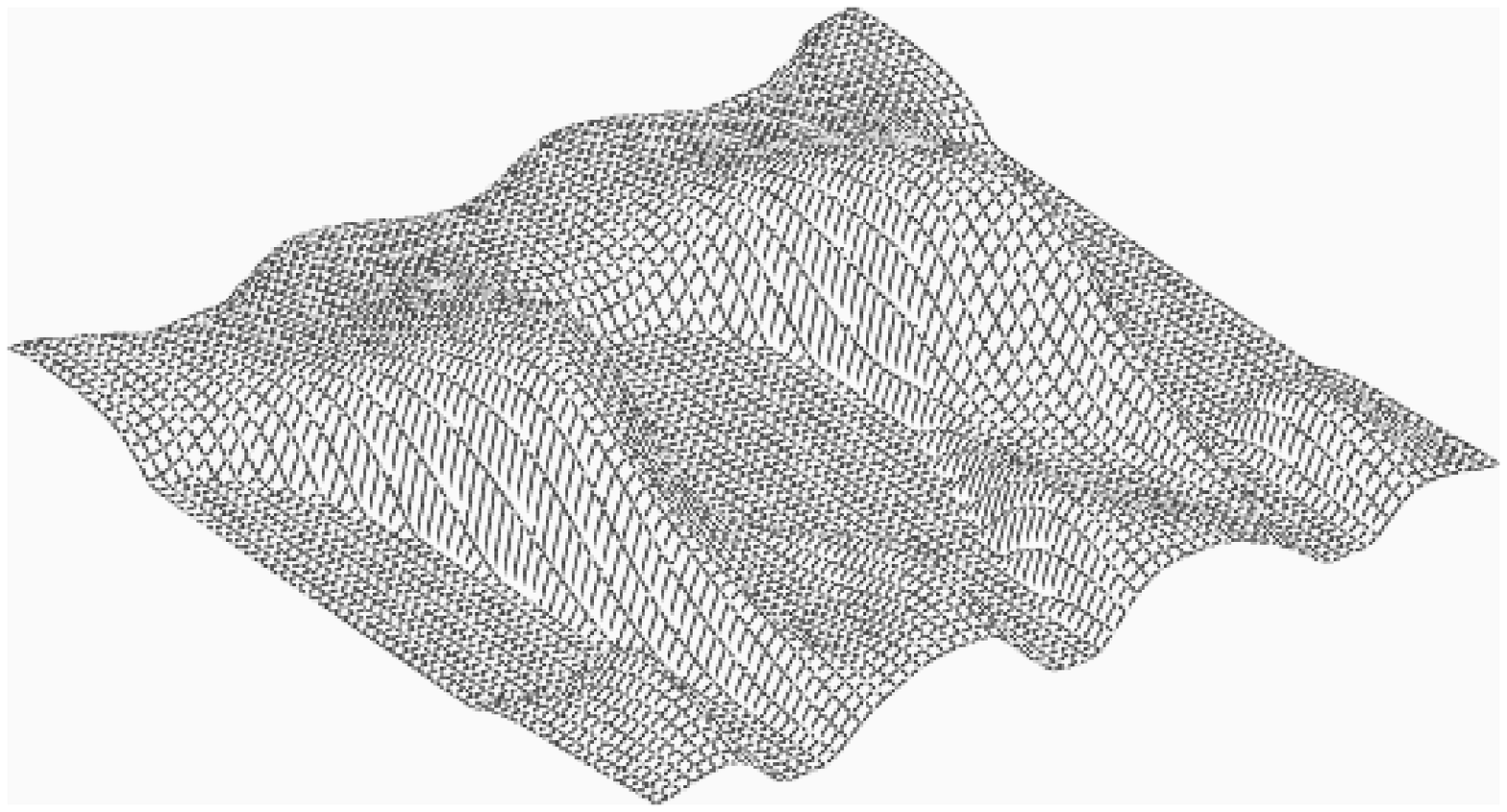}}

\caption{(a) Positive and (b) negative
part of the correlation $C_{z}(\tau_{11})$ for
the nonintegrable system.}
\label{fig 6}
\end{figure}

Four surfaces of section in the energy range $[8,9]$ are shown in
Fig.~(\ref{fig 5}).  The classical dynamics does not vary significantly in
this energy range and exhibits a mixed phase space.  The two resonances
have overlapped, generating stochastic regions in the phase space.
Yet, it is still possible to identify the stable fixed points
corresponding to the two resonances.  Figure~(\ref{fig 6}) shows the
positive and negative components of the correlation function
$C_{z}(\tau_{11})$.  As compared to the integrable case in Fig.~(\ref{fig
3}), there are many common features including the large positive values
at the stable fixed point and the associated dip.  This indicates that the
linear parametric motion persists at these energies and coupling
strengths. There are, however, important differences between the mixed
phase space and the integrable cases. In particular the negative regions
of the correlation in Fig.~(\ref{fig 6}b) show a lot more structure.  The
dip associated with the unstable fixed point is hardly noticeable.  This
is consistent with the destruction of the separatrix due to the overlap of
the two primary resonances.  Notice that the negative values of the
correlation are mainly associated with the chaotic regions of the phase
space as opposed to the positive values shown in Fig.~(\ref{fig 6}a). The
nonuniformity of the negative $C_{z}(\tau_{11})$ implies that there exist
eigenstates in the energy range which are affected strongly by the $1:1$
perturbation and which share common localization features.
Figure~(\ref{fig 6}a) clearly indicates the existence of strongly
localized states close to the center of the $1:1$ resonance channel. 
Roughly speaking, although in the limit of strongly chaotic dynamics, the
chaotic region should show no structure, the chaotic region in a mixed
phase space system has a great deal of structure related to cantori or
broken separatrices, and ``stickiness'' whereby trajectories become
trapped near tiny regular regions.  Transport is not rapid and uniform,
and the localizing effects on the eigenstates show up in the correlation
function.  In order for more details of this structure to be visible in
Fig.~(\ref{fig 6}a), a much higher density of states (smaller value of
$\hbar$) would have to be used.  

\begin{figure}
\subfigure[]
{\includegraphics*[width=3.5in,height=1.5in]{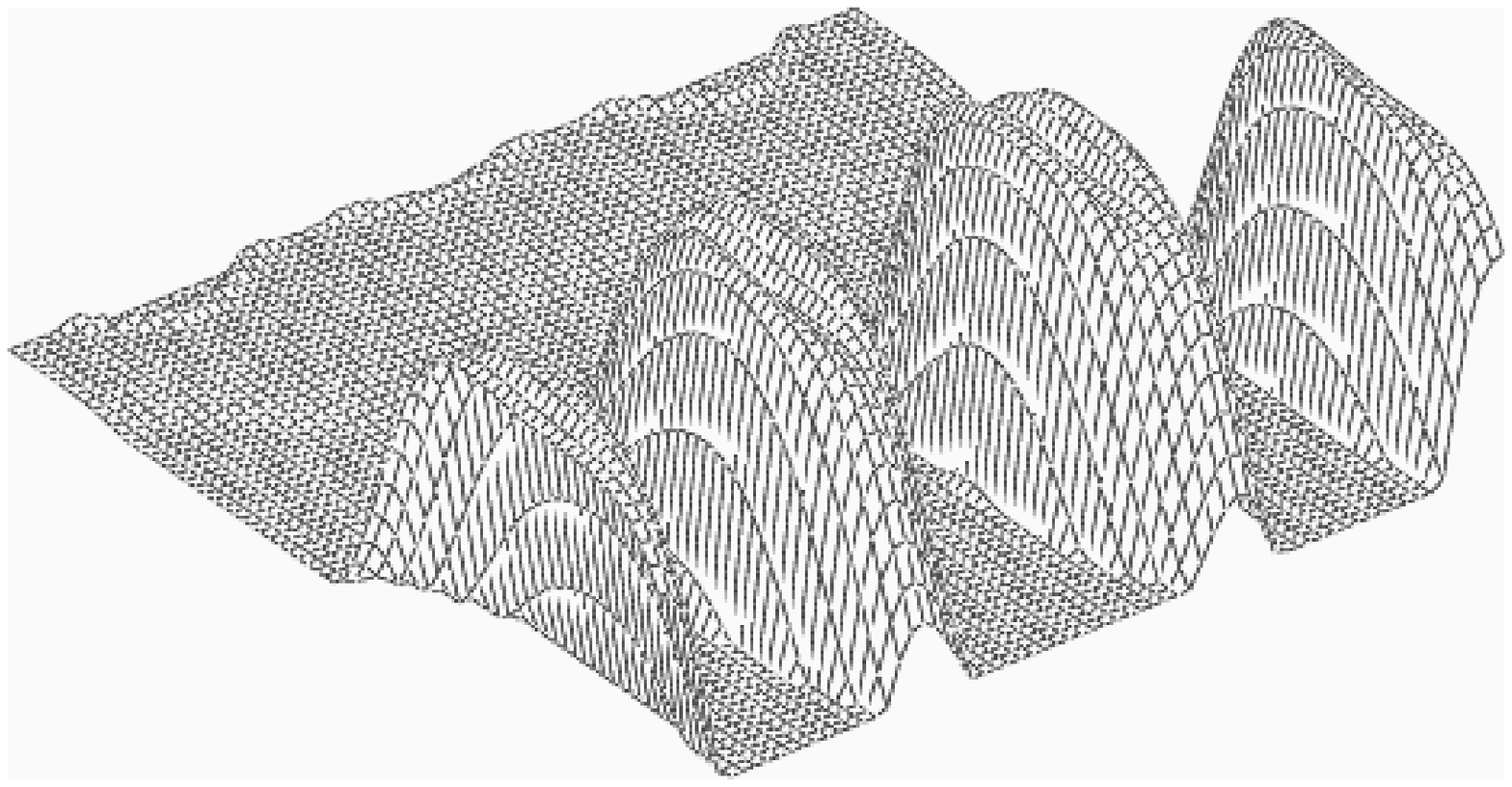}}

\subfigure[]
{\includegraphics*[width=3.5in,height=1.5in]{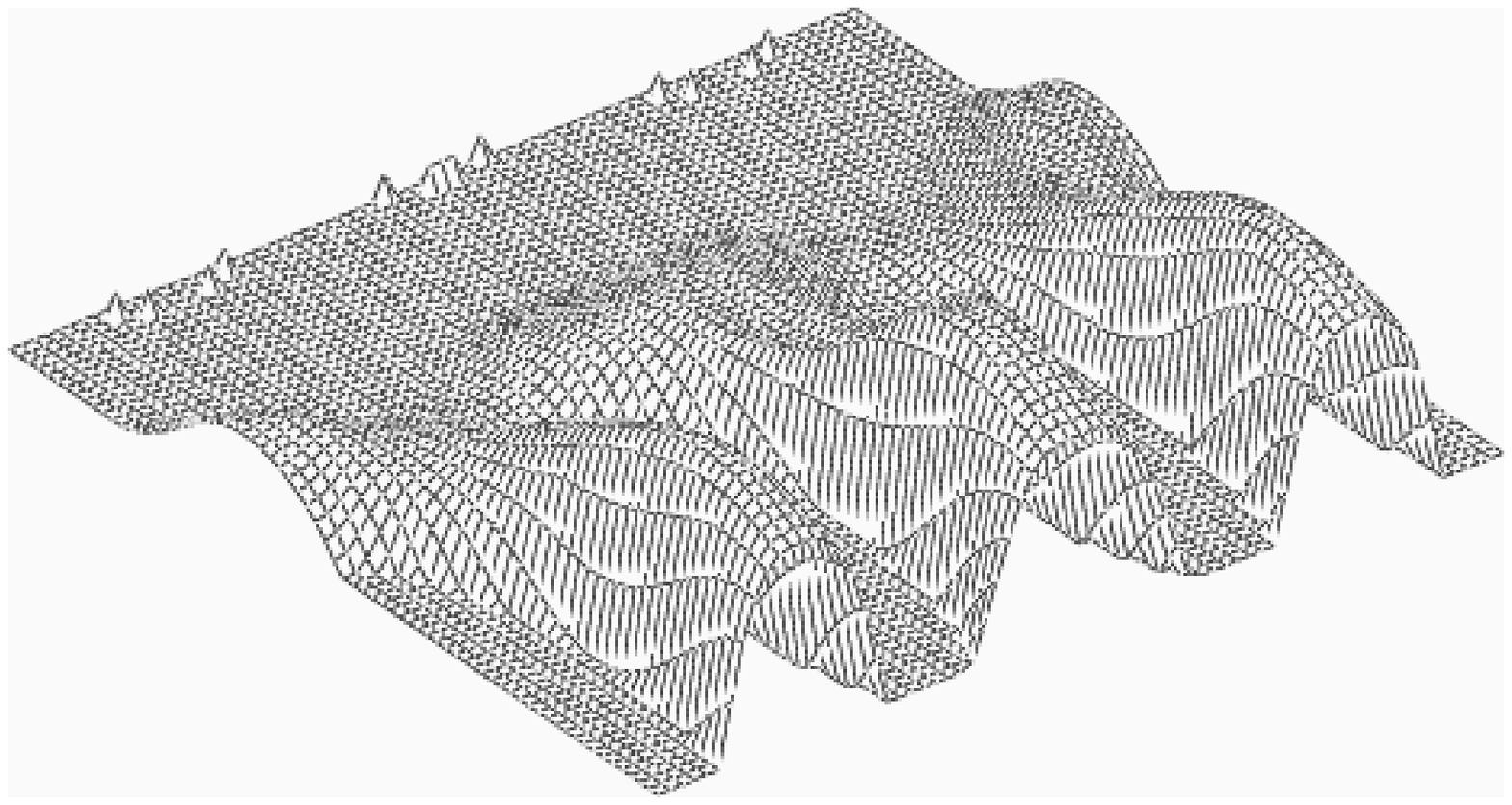}}
\caption{(a) Positive and (b) negative part 
of the correlation $C_{z}(\tau_{21})$ for
the nonintegrable system.}
\label{fig 7}
\end{figure}

Figure~(\ref{fig 7}) shows the analogus results for the correlation
$C_{z}(\tau_{21})$.  The results are similar to that of the $1:1$ case.
Parametric variations in $\tau_{21}$ highlight the $2:1$ region of the
phase space.  Note that it is quite possible to choose a state $|z\rangle$
such that both the correlations are negligibly small.  This implies that
the chosen $|z\rangle$ is not strongly affected by either of the
resonances.  Thus, if a coherent state $|z\rangle$ lies in a region that
is strongly affected by a certain perturbation, then the associated
correlation will highlight the relevant part of the phase space. 

\begin{figure}
\includegraphics*[width=3.0in,height=3.0in]{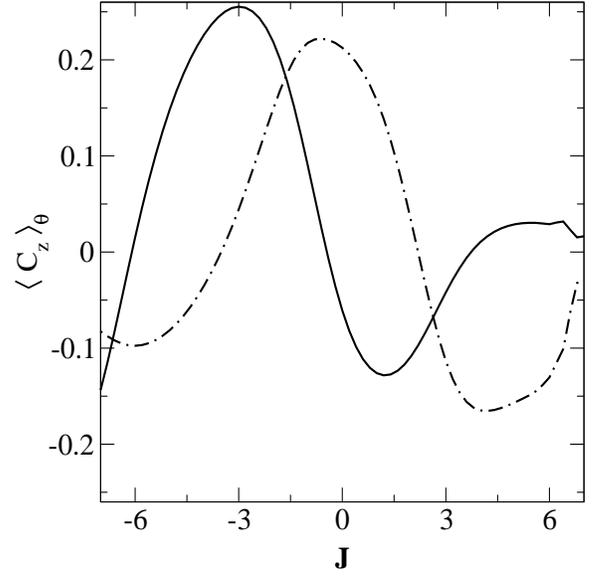}
\caption{Angle averaged correlations in the nonintegrable case. The solid
line corresponds to the $2:1$ case and the dot-dashed line corresponds to the
$1:1$ correlation. The maximum in each case is very close to the
corresponding resonant center $J$ values. For the parameters chosen in
this work $J^{r}_{2:1} \approx -3.75$ while $J^{r}_{1:1}$ is
consistent with the maximum of the averaged correlation. }
\label{fig 8}
\end{figure}

The model system considered in this paper has two degrees of freedom.
Consequently it is possible to obtain detailed information about the
phase space structure and localization features of eigenstates.
In three or more degrees of freedom systems the standard approach
of surface of sections and phase space distribution functions
cannot be employed for well known reasons\cite{licht}.  The results in
this section demonstrate that the correlation function highlights the
various classical structures (periodic orbits, tori, etc.) in the phase
space.  Nevertheless, the correlation function is not subject to the
usual constraints that have to be satisfied by a surface of section.
This suggests that it is possible to compute the correlation over various
reduced dimensionality spaces and gain insights into the system.  For
instance, the correlation could be averaged over the angle variables, and
obtain an average correlation as a function of the actions alone.  As a
simple example in Fig.~(\ref{fig 8}), we show the angle averaged
correlations $\langle C_{z}(\tau)\rangle_{\bm \theta}$ corresponding to
the results in Fig.~(\ref{fig 6},\ref{fig 7}).  The peaks of the averaged
correlation function are very close to the center of the respective
resonance zones. 
Hence,
$\langle C_{z}(\tau)\rangle_{\bm \theta}$ is providing information on the
local resonance structure associated with a particular $|z\rangle$ which
is linked\cite{rev} experimentally to the gateway states.

\begin{figure}
\includegraphics*[width=3.0in,height=3.0in]{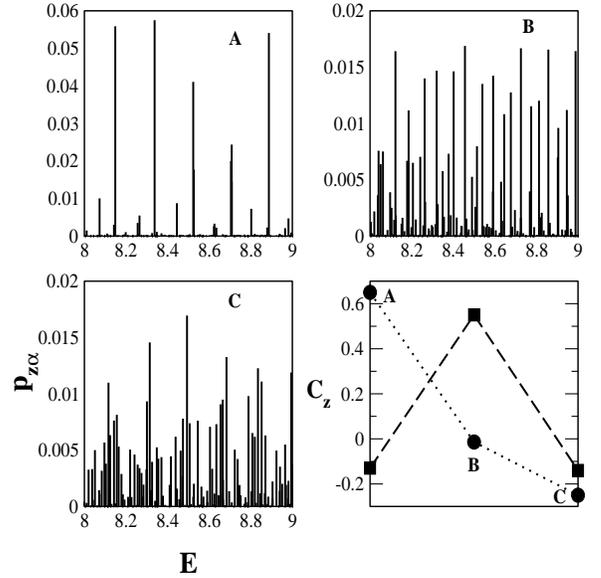}
\caption{Spectra associated with three different coherent states and the
correlation functions. $\hbar = 1/3$ in this
figure. (A) $|z_{A}\rangle$ centered close to the
$1:1$ stable fixed point.
(B) $|z_{B}\rangle$ centered close to the $2:1$ stable fixed point.
(C) $|z_{C}\rangle$ centered in the chaotic region. 
The bottom right panel shows the correlations for the three coherent
states with circles denoting $C_{z}(\tau_{11})$ and squares denoting
$C_{z}(\tau_{21})$. The lines are drawn as a guide to the eye.}
\label{fig 9}
\end{figure}

As metioned in the introduction, experimentally there is access to very
few bright states in a particular energy range.  One of the important
questions, given the bright state and its associated spectrum, concerns
the relative effect of the various perturbations in a certain energy
range. It is of particular interest to demonstrate how the above
question can be addressed with the overlap intensity-level velocity
correlation function.  In order to determine the usefulness of the
correlation in this regard, we select three different bright states
$|z_{A}\rangle, |z_{B}\rangle$ and, $|z_{C}\rangle$.  The first two have
been introduced in the previous subsection, and the third one
$|z_{C}\rangle$ is centered at $(J,\psi) = (0.6,-1.5)$ corresponding to
the chaotic region of the phase space.  The spectra associated with the
three states are shown in Fig.~(\ref{fig 9}).  In order to have a
higher density of states, the value of $\hbar = 1/3$.  For the chosen
value of $\hbar$, there are $250$ states in the energy range shown in
Fig.~(\ref{fig 9}).  The correlations $C_{z}(\tau_{11})$ and
$C_{z}(\tau_{21})$ associated with each of the spectra are also shown.
It is clear that for the bright state $|z_{A}\rangle$ the $1:1$
perturbation has a much larger effect as compared to the $2:1$
perturbation.  The roles are reversed in the case of the bright state
$|z_{B}\rangle$.  In the case of the state $|z_{C}\rangle$, both of the
correlations are negative and small.  However, the correlations are not
insignificant since, as a rough estimate, a RMT-like result would give
$C_{z} \sim 0 \pm 0.063$.  Moreover, the correlation is almost two times
larger for the $1:1$ perturbation as compared to the $2:1$ perturbation. 
>From a time-dependent point of view, recurrences in the survival
amplitude $\langle z|z(t)\rangle$ are connected to classical structures
in the phase space.  The correlation function is providing clues as to the
precise nature of the classical phase space structures which are
dictating the dynamics of the bright state of interest. 

Finally, a closer look at the correlations shown in Figs.~(\ref{fig 6},
\ref{fig 7}), compared to their integrable counterparts, indicates the
spreading of the correlation due to the destruction of barriers in the
phase space.  The classical surfaces of sections shown in Fig.~(\ref{fig
5}) have been computed up to times significantly larger than the
heisenberg time $\tau_{H}$.  The fact that the correlations are indicating
``amplitude transport" through broken KAM tori or separatrices gives
a unique probe of the transport.  Moreover, the uneven spreading of the
correlations suggests the existence of partial barriers.  This
observation raises the possibility of understanding transport phenomena
in the phase space via the correlation function.  Further work is
warranted to establish what information is contained, and how to extract
it from the correlation behaviors in the chaotic regions of mixed
phase space systems.   

\section{Summary and Conclusions}

In this paper we have explored the possibility of studying IVR through the
overlap intensity-level velocity correlator.  A semiclassical theory
for the correlator has been developed in the case that the underlying
classical dynamics is integrable.  Combined with the earlier
work\cite{cllt} providing the semiclassical theory for chaotic systems,
the correlator is an ideal candidate for a detailed study of the
classical-quantum correspondence for a general system.  Application to an
effective spectroscopic Hamiltonian clearly supports our point of view. 
Although the model Hamiltonian used in the present study is two
dimensional, it is evident that the correlator is well suited for
multidimensional systems with a high density of states.  It was
demonstrated that choosing different parameters, corresponding to the
different perturbations, highlighted the appropriate parts of the phase
space displaying prominent localization features.  The fact that these
localization features are often connected to classical structures like
tori, resonance zones, and periodic orbits provides an avenue to
understand the relative importance of the various perturbations in a
given energy range.  From the IVR perspective this information is
intimately connected to the dynamics of an initial bright state of
interest.  If the dynamics of a particular bright state is influenced
strongly by specific classical structures, then the correlation will
invariably reveal the relevant terms in the molecular Hamiltonian which
play an important role.

The manner in which we have applied the technique in this work is useful
in situations where one has an effective spectroscopic Hamiltonian at
hand.  For a specific molecule, the coupling strengths are fixed and
experimentally one cannot vary the perturbation strengths in a direct
fashion.  However, the standard practice of isotopic
labeling/substitution achieves this in an indirect, albeit
uncontrollable, way since such studies\cite{dark} do vary the nature of
the dark states relative to the bright state.  A more direct possibility
would be to use various external fields or parameters to generate the
level motions; but then the connection between the fields and the possible
Hamiltonian terms would need to be explored.  From a theoretical
standpoint, the computation of the intensities requires the eigenstates
which can be a daunting task for molecules with a large number of coupled
modes.  Significant advances in computational methods in recent years is
certainly encouraging\cite{rev5}.  However, under some circumstances, a
hybrid approach of correlating the experimental intensities with the
computed level velocities could be tried.  Another possibility is to adapt
the MFD approach\cite{mfd2} where intensities are related to the
parametric derivatives of the eigenvalues.

Throughout this paper we have assumed the availability of an eigenstate
resolved spectrum for the system.  At a coarser level, corresponding to
shorter time scales, there is the dispersed fluorescence spectrum which
contains key information regarding IVR and approximately conserved
quantities.  It would be interesting to have an explicit time-dependent
view of the IVR process {\it i.e.,} deciding the perturbations which
dictate the dynamics of a bright state over different time scales.  For
this purpose, it is appropriate to work with the finite time or smoothed
strength function\cite{strength} defined as:
\begin{equation}
S^{(T)}_{z}(E,\tau) = \frac{1}{2 \pi \hbar} \int_{-T}^{T} dt
e^{i E t/\hbar} \langle z|e^{-i \widehat{H}(\tau)t/\hbar}|z\rangle 
\end{equation}
There is a problem with using the finite time strength function since the
concept of eigenstates and hence that of level velocities is not
meaningful.  We are currently exploring the correct finite time
analog of the correlator. 

As stated earlier there are many aspects of the intensity-level velocity
correlator which warrant further studies.  Foremost among them are phase
space transport, spectroscopy in external fields and  quantum versus
classical routes to localization.  Efforts are currently underway to
systematically approach and understand these features in molecular
systems.

\section{Acknowledgments}
This work was supported by grants from the National Science Foundation,
PHY-0098027, and the Office of Naval Research, N00014-98-1-0079. SK thanks
Tapas Chakraborty for helpful discussions and the Department of Science
and Technology and, Council for Scientific and Industrial Research, India
for financial support.


\begin{thebibliography}{}
\bibitem{rev}{For a recent review see 
M. Gruebele, Adv. Chem. Phys. {\bf 114}, 193 (2000)
and references therein.}
\bibitem{rev1}{P. M. Felker and A. H. Zewail, Adv. Chem. Phys. {\bf 70},
265 (1988).}
\bibitem{rev3}{K. K. Lehmann, G. Scoles, and B. H. Pate, Annu. Rev.
Phys. Chem. {\bf 45}, 241 (1994).}
\bibitem{rev4}{D. J. Nesbitt and R. W. Field, J. Phys. Chem. {\bf 100},
12735 (1996).}
\bibitem{rev5}{M. Gruebele and R. Bigwood, Int. Rev. Phys. Chem.
{\bf 17}, 91 (1998).}
\bibitem{rev6}{See for example {\em Molecular Dynamics and
Spectroscopy by Stimulated Emission Pumping}, H. -L. Dai and R. W. Field
Eds., World Scientific, 1995.}
\bibitem{rev7}{A. H. Zewail, J. Phys. Chem. {\bf 100}, 18666 (1996).}
\bibitem{rev8}{See articles in {\em Lase Spectroscopy of Highly
Vibrationally Excited Molecules}, V. S. Letokhov (ed.), Adam Hilger, Bristol
(1989).}
\bibitem{rev2}{T. Uzer, Phys. Rep. {\bf 199}, 73 (1991).}
\bibitem{rev9}{K. S. J. Nordholm and S. A. Rice, J. Chem. Phys. {\bf 61},
203 (1974).}
\bibitem{rev10}{K. G. Kay, J. Chem. Phys. {\bf 72}, 5955 (1980).}
\bibitem{benzene}{ A. Callegari, H. K. Srivatsava, 
U. Merker, K. K. Lehmann, G. Scoles,
and M. J. Davis, J. Chem. Phys. {\bf 106}, 432 (1997).}
\bibitem{hcp1}{H. Ishikawa, R. W. Field, S. C. Farantos, M. Joyeux,
J. Koput, C. Beck, and R. Schinke, Annu. Rev. Phys. Chem. {\bf 50},
443 (1999).}
\bibitem{hocl1}{R. Jost, M. Joyeux, S. Skokov, and J. Bowman,
J. Chem. Phys. {\bf 111}, 6807 (1999).}
\bibitem{ethy}{See for example, M. P. Jacobson, R. J. Silbey, and
R. W. Field, J. Chem. Phys. {\bf 110}, 845 (1999).}
\bibitem{water}{J. E. Baggot, Mol. Phys. {\bf 65}, 739 (1988).}
\bibitem{qu1}{See M. Quack, Annu. Rev. Phys. Chem. {\bf 41},
839 (1990) for a review about the construction of effective
spectroscopic Hamiltonians.}
\bibitem{qu2}{A. Beil, D. Luckhaus, M. Quack, and J. Stohner,
Ber. Bunsenges. Phys. Chem. {\bf 101}, 311 (1997).}
\bibitem{ethy1}{M. P. Jacobson, C. Jung, H. S. Taylor, and R. W. Field,
J. Chem. Phys. {\bf 111}, 600 (1999).}
\bibitem{water1}{S. Keshavamurthy and G. S. Ezra, J. Chem. Phys. {\bf 107},
156 (1997).}
\bibitem{qs0}{See, M. P. Jacobson and M. S. Child, J. Chem. Phys. {\bf 114},
250 (2001), for a discussion on the validity of effective spectroscopic
Hamiltonians in systems that undergo isomerization.}
\bibitem{qs1}{E. L. Sibert III, W. P. Reinhardt, and J. T. Hynes,
J. Chem. Phys. {\bf 81}, 1115 (1984).}
\bibitem{qs2}{Y. M. Engel and R. D. Levine, Chem. Phys. Lett.
{\bf 164}, 270 (1989).}
\bibitem{qs3}{M. E. Kellman, Annu. Rev. Phys. Chem. {\bf 46},
395 (1995) and references therein.}
\bibitem{qs4}{B. Zhilinskii, Spectrochim. Acta A {\bf 52}, 881 (1996) and
references therein.}
\bibitem{qs5}{G. S. Ezra, Adv. Class. Traj. Meth. {\bf 3}, 35 (1998) and
references therein.}
\bibitem{qs6}{C. C. Martens and W. P. Reinhardt, J. Chem. Phys. {\bf 93},
5621 (1990).}
\bibitem{qs7}{C. Jung, E. Ziemniak, and H. S. Taylor, J. Chem. Phys.
{\bf 115}, 2499 (2001).}
\bibitem{qs8}{M. Lewerenz and M. Quack, J. Chem. Phys. {\bf 88},
5408 (1988).}
\bibitem{3dwork}{C. C. Martens, J. Stat. Phys. {\bf 68}, 207 (1992).}
\bibitem{husimi}{K. Husimi, Proc. Phys. Math. Japan, {\bf 22},
264 (1940).}
\bibitem{meth}{O. V. Boyarkin, L. Lubich, R. D. F. Settle, D. S. Perry,
and T. R. Rizzo, J. Chem. Phys. {\bf 107}, 8409 (1997).}
\bibitem{mork}{S. W. Mork, C. C. Miller, and L. A. Philips,
J. Chem. Phys. {\bf 97}, 2971 (1992).}
\bibitem{tier1}{R. E. Wyatt, C. Iung, and C. Leforestier, J. Chem. Phys.
{\bf 97}, 3477 (1992).}
\bibitem{tier2}{Y. Zhang and R. A. Marcus, J. Chem. Phys. {\bf 96},
6065 (1992).}
\bibitem{tier3}{A. A. Stuchebrukhov and R. A. Marcus, J. Chem. Phys.
{\bf 98}, 6044 (1993).}
\bibitem{tier4}{R. Bigwood and M. Gruebele, Chem. Phys. Lett.
{\bf 235}, 604 (1995).}
\bibitem{watercpl}{S. Keshavamurthy and G. S. Ezra, Chem. Phys. Lett.
{\bf 259}, 81 (1996).}
\bibitem{dyntunn}{M. J. Davis and E. J. Heller, J. Chem. Phys. {\bf 75},
246 (1981).}
\bibitem{chaotunn}{S. Tomsovic and D. Ullmo, Phys. Rev. E {\bf 50}, 145
(1994).}
\bibitem{irir1}{B. H. Pate, K. K. Lehmann, and G. Scoles, J. Chem. Phys.
{\bf 95}, 3891 (1991).}
\bibitem{irir2}{J. Go and D. S. Perry, J. Chem. Phys. {\bf 97},
6994 (1992).}
\bibitem{eros}{G. T. Fraser and B. H. Pate, J. Chem. Phys. {\bf 98},
2477 (1993).}
\bibitem{eros2}{G. T. Fraser and B. H. Pate, J. Chem. Phys. {\bf 100},
6210 (1994).}
\bibitem{hier1}{M. J. Davis, J. Chem. Phys. {\bf 98}, 2614 (1993).}
\bibitem{hier2}{M. J. Davis, Int. Rev. Phys. Chem. {\bf 14},
15 (1995).}
\bibitem{propyne}{J. E. Gambogi, E. R. Th. Kerstel, K. K. Lehmann, 
and G. Scoles, J. Chem. Phys. {\bf 100}, 2612 (1994).}
\bibitem{woly}{S. Schofield and P. G. Wolynes, J. Chem. Phys. {\bf 98},
1123 (1993).}
\bibitem{sri}{S. Keshavamurthy, Chem. Phys. Lett. {\bf 300}, 281 (1999).}
\bibitem{grueb1}{M. Gruebele, Proc. Natl. Acad. Sci. USA {\bf 95},
5965 (1998).}
\bibitem{grueb2}{R. Bigwood, M. Gruebele, D. M. Leitner, and
P. G. Wolynes, Proc. Natl. Acad. Sci. USA {\bf 95}, 5960 (1998).}
\bibitem{mfd1}{M. Gruebele, J. Chem. Phys. {\bf 104}, 2453 (1996).}
\bibitem{mfd2}{M. Gruebele, J. Phys. Chem. {\bf 100}, 12178 (1996).}
\bibitem{weijort}{Y. Weissman and J. Jortner, J. Chem. Phys. {\bf 77},
1486 (1982).}
\bibitem{billiard}{S. W. McDonald, Ph. D. dissertation, University
of California, Berkeley, Lawerence Berkeley Laboratory Report No.
LBL-14837 (1983).}
\bibitem{radons}{G. Radons, T. Geisel, and J. Rubner, Adv. Chem. Phys.
{\bf LXXIII}, 891 (1989).}
\bibitem{bsep}{O. Bohigas, S. Tomsovic, and D. Ullmo, Phys. Rep. {\bf 223},
43 (1993).}
\bibitem{cantor}{R. Ketzmerick, G. Petschel, and T. Geisel,
Phys. Rev. Lett. {\bf 69}, 695 (1992).}
\bibitem{diffus}{S. Fishman, D. R. Grempel, and R. E. Prange,
Phys. Rev. A {\bf 36}, 289 (1987).}
\bibitem{scar}{E. J. Heller, Phys. Rev. Lett. {\bf 53}, 1515 (1984).}
\bibitem{bog} E. B. Bogomolny, Physica D {\bf 31}, 169 (1988).
\bibitem{kaplan} L. Kaplan and E. J. Heller, Ann. Phys. NY {\bf 264}, 171
(1998).
\bibitem{bgs} O. Bohigas, M.-J. Giannoni, and C. Schmit, Phys. Rev. Lett.
{\bf 52} 1 (1984).
\bibitem{stechel} E. B. Stechel and E. J. Heller, Ann. Rev. Phys. Chem.
{\bf 35}, 563 (1984).
\bibitem{stprl}{S. Tomsovic, Phys. Rev. Lett. {\bf 77}, 4158 (1996).}
\bibitem{cllt}{N. R. Cerruti, A. Lakshminarayan, J. H. Lefebvre, and
S. Tomsovic, Phys. Rev. E {\bf 63}, 016208 (2001).}
\bibitem{lct}{A. Lakshminarayan, N. R. Cerruti, and S. Tomsovic,
Phys. Rev. E {\bf 63}, 016209 (2001).}
\bibitem{rmt}{See for example, T. Guhr, A. M\"{u}ller-Groeling, and
H. A. Weidenm\"{u}ller, Phys. Rep. {\bf 299}, 189 (1998), and
references therein.}
\bibitem{finger}{ S. Keshavamurthy, J. Phys. Chem. A {\bf 105}, 2668 (2001).}
\bibitem{parisri}{A. Semparithi, V. Charulatha and S. Keshavamurthy, 
J. Chem. Phys. (to
be submitted).}
\bibitem{nygard}{See J. Nygard, {\em Quantum Chaos of the NO$_{2}$
Molecule in High Magnetic Fields}, Master's Thesis, University of
Copenhagen (1996).}
\bibitem{heisen}{W. Heisenberg, Z. Phys. {\bf 33}, 879 (1925). Translated
in {\em Sources of Quantum Mechanics}, B. L. Van der Waerden (Ed.), Dover,
Minneola, NY (1967).}
\bibitem{strength}{See E. J. Heller, in {\em Chaos and Quantum Physics},
eds. M. J. Giannoni, A. Voros, and J. Zinn-Justin (Elsevier,
Amsterdam, 1991).}
\bibitem{licht}{A. J. Lichtenberg and M. A. Lieberman, {\em Regular and
Stochastic Motion}, Springer-Verlag, NY (1983).}
\bibitem{poly}{L. E. Fried and G. S. Ezra, J. Chem. Phys. {\bf 86},
6270 (1987).}
\bibitem{later}{N. R. Cerruti, S. Keshavamurthy, and S. Tomsovic,
to be submitted.}
\bibitem{joy}{M. Joyeux, Chem. Phys. {\bf 185}, 263 (1994).}
\bibitem{amod}{A. M. Ozorio de Almeida, {\em Hamiltonian Systems: Chaos
and Quantization}, Cambridge University Press (1988).}
\bibitem{ugt} D. Ullmo, M. Grinberg, and S. Tomsovic, Phys. Rev. E {\bf
54}, 136 (1996).
\bibitem{keating}{M. V. Berry and J. P. Keating, J. Phys. A: Math. Gen.
{\bf 27}, 6167 (1994).}
\bibitem{leboeuf}{P. Leboeuf and M. Sieber, Phys. Rev. E {\bf 60}, 3969 
(1999).}
\bibitem{bt}{M. V. Berry and M. Tabor, J. Phys. A: Math. Gen.
{\bf 10}, 371 (1977).}
\bibitem{note}{The validity of the expression can be explicitly checked
in the case of a pendulum Hamiltonian. In addition note that
$\partial S_{\bf M}/\partial \tau = T_{\bf M}\, \partial E/\partial \tau$.}
\bibitem{hozsum}{J. H. Hannay and A. M. Ozorio de Almeida,
J. Phys. A: Math. Gen. {\bf 17}, 3429 (1984).}
\bibitem{takhas}{T. Takami and H. Hasegawa, Phys. Rev. Lett. {\bf 68},
419 (1992).}
\bibitem{error}{This scaling of the variance of the intensities differs
from our previous publication (Ref.~\onlinecite{cllt}), which contained
an error in the derivation of the average intensities and sum rule.}
\bibitem{dark}{See for example, O. V. Boyarkin, T. R. Rizzo, and
D. S. Perry, J. Chem. Phys. {\bf 110}, 11346 (1999).}

\end{thebibliography}
\end{document}